# Adaptive frequency-based modeling of whole-brain oscillations: Predicting regional vulnerability and hazardousness rates

Neda Kaboodvand[1,*], Martijn P van den Heuvel[2], Peter Fransson[1]

[1]Department of Clinical Neuroscience, Karolinska Institutet, Stockholm, Sweden.

[2]Connectome Lab, Department of Complex Trait Genetics, CNCR, VU Amsterdam, The Netherlands.

*Corresponding author

Department of Clinical Neuroscience

Karolinska Institutet

Nobels väg 9

SE-171 77 Stockholm, SWEDEN

E-mail: Neda.Kaboodvand@ki.se

**Abstract**

Whole-brain computational modeling based on structural connectivity has shown great promise in successfully simulating fMRI BOLD signals with temporal co-activation patterns that are highly similar to empirical functional connectivity patterns during resting state. Importantly, previous studies have shown that spontaneous fluctuations in co-activation patterns of distributed brain regions have an inherent dynamic nature with regard to the frequency spectrum of intrinsic brain oscillations. In this modeling study, we introduced frequency dynamics into a system of coupled oscillators, where each oscillator represents the local mean-field model of a brain region. We first showed that the collective behavior of interacting oscillators reproduces previously shown features of brain dynamics. Second, we examined the effect of simulated lesions in gray matter by applying an *in silico* perturbation protocol to the brain model. We present a new approach to map the effects of vulnerability in brain networks and introduce a measure of regional hazardousness based on mapping of the degree of divergence in a feature space.

**Author summary**

Computational modeling of the brain enables us to test different hypotheses without any experimental complication and it provides us with a platform for improving our understanding of different brain mechanisms. In this study, we proposed a new macroscopic computational model of the brain oscillations for resting-state fMRI. Optimizing model parameters using empirical data was performed based on several measures of functional connectivity and instantaneous coherence. We simulated the effect of malfunction in a brain region by changing that region's dynamics to evoke noisy behavior. Together with presenting a new paradigm for local vulnerability mapping in the brain connectome, we evaluated the hazard rate induced after perturbing a brain region by measuring divergence of the perturbed model from the original model in feature space. The analysis of hazard rates induced by primary failures of individual brain regions provides relevant insights not only into the size of the damage inflicted on the connectome by a particular failure, but also into the potential origins of disease. Furthermore, we proposed a spatial brain map that is associated with the regional hazardousness rates, which is in good agreement with the known pathophysiologic roles of malfunction in different functional subsystems in the brain.

1. **Introduction**

The human connectome is a complex network that is made up of interactive systems encompassing a large number of regions. Regions communicate with each other in order to share and process information providing the structural and functional basis for complex cognitive processes. Understanding the network architecture of human brain in the context of its robustness and the integration between different subnetworks (known as functional systems) has received increased attention in system-level network neuroscience. However, we still have limited knowledge of the emergence of brain dynamics from the underlying anatomy. So far, there is evidence suggesting that the large-scale structural connectome of the brain constrains the strength and persistence of resting state functional connectivity (FC) (C J Honey et al., 2009), with significant contributions of structural connections for the integrated state (Fukushima, Betzel, He, van den Heuvel, et al., 2018). An impaired structural connectome may lead to disrupted FC which contributes to neurodegenerative diseases (for example, see (Griffa, Baumann, Thiran, & Hagmann, 2013)). The interplay between the brain's structure and dynamics underlies all brain functions spanning from consciousness and perception to learning, memory and movement (Deco, Jirsa, Robinson, Breakspear, & Friston, 2008). Moreover, the relationship between the structural backbone and the dynamics of brain activity is believed to play an important role for surviving network communication failures and attacks (Barabási, 2016). In the past decade we have witnessed great progress towards the systematic modeling of the neural network dynamics (Breakspear, Heitmann, & Daffertshofer, 2010; Cabral, Kringelbach, & Deco, 2014; Fink, 2018). Large-scale computational models are uniquely suited to address difficult questions related to the role of the brain's structural network in shaping measures of FC averaged across longer time-scales (Cabral et al., 2014; Ritter, Schirner, McIntosh, & Jirsa, 2013). Additionally, we can study the emergence of complex brain dynamics (Roberts et al., 2018), through time-varying analyses of functional coherence.

Several resting-state studies have shown that there exist spontaneous fluctuations in co-activation patterns of distributed brain regions (Hutchison et al., 2013; Thompson, Brantefors, & Fransson, 2017; Zalesky, Fornito, Cocchi, Gollo, & Breakspear, 2014), which gives rise to an efficient information exchange while minimizing metabolic expenditure (Zalesky et al., 2014). Furthermore, spectral analysis of brain fluctuations has disclosed valuable information about the underlying sources of time-varying connectivity of brain regions (C. Chang & Glover, 2010; Ries et al., 2018; Yaesoubi, Allen, Miller, & Calhoun, 2015). These studies suggest that the frequency spectrum of intrinsic oscillations of the brain has a time-varying nature.

In this study, we present a theoretical framework for modeling large-scale brain dynamics based on the theory of dynamical systems. Dynamical systems theory is a mathematical framework used to describe the behavior of the complex dynamical systems (i.e., systems that evolve in time), usually by a set of differential equations (Steven H. Strogatz, 2018). The proposed model is used to show that important questions related to the prediction of perturbation patterns can be tackled in an accessible and useful way. Furthermore, we show that our model can be employed to simulate perturbations in different brain regions to assess the vulnerability and hazardousness of individual connections and brain regions.

To do this, we start by constructing a macroscopic computational model of the brain where local brain regions, each modeled by a local mean-field model, are interacting through the structural brain network architecture (Breakspear, 2017; Cabral et al., 2014; Deco & Jirsa, 2012; Deco, Jirsa, McIntosh, Sporns, & Kötter, 2009; Gollo, Zalesky, Hutchison, van den Heuvel, & Breakspear, 2015; C J Honey et al., 2009; Christopher J Honey, Kötter, Breakspear, & Sporns, 2007). We then suggest that the local dynamics of each brain area can be described by a modified Stuart-Landau equation. Subsequently, we show that simulated BOLD signals are forming FC patterns that are similar to the FC patterns obtained from measured BOLD signals. This similarity was found to be valid both in terms of strength of FC and in the form of the establishment of communities of brain regions as shown in previous resting-state networks studies (Yeo et al., 2011). Additionally, we show that temporal structure of simulated BOLD signals is highly similar to the fluctuations of empirical BOLD signals.

Of note, by perturbing different brain regions in our model and measuring how the system responds to the induced failures, we can obtain information on the underlying association between structure and dynamics in the brain. Previous studies have simulated the effects of brain lesions on both local and global levels of network activity, for example by removing individual connections (edges) or all connections of an individual region from the network (Aerts, Fias, Caeyenberghs, & Marinazzo, 2016; Cabral, Hugues, Kringelbach, & Deco, 2012; Deco, Van Hartevelt, Fernandes, Stevner, & Kringelbach, 2017; Váša et al., 2015). However, applying models with Hopf bifurcation are shown to be particularly well-suited for *in silico* perturbation studies (Deco et al., 2018; Saenger et al., 2017). Therefore, in this study we used an *in silico* perturbation protocol by applying bifurcation-induced shifts in the dynamical regime of each individual brain region. Regime refers to the characteristic behavior of a dynamical system and regime shifts are sudden, large and persistent changes in the function of the system as a result of some external source of disturbance (Folke et al., 2004; Holling, 1973; Scheffer, Carpenter, Foley, Folke, & Walker, 2001). Static as well as dynamic measures of FC patterns were investigated for different *in silico* failures. This analysis was followed up by applying the representational similarity analysis (RSA) framework to investigate how a targeted brain region's failure contributed to an increased distance between the perturbed connectome versus the healthy connectome in RSA space. Further, we suggest that the aforementioned distance in turn provides useful information about the degree of regional hazardousness. Moreover, we propose that our approach to modeling brain dynamics is helpful to understand the diversity of fragility for brain regions in the connectome with regard to injuries and disease. We suggest that investigating perturbation maps which are aggregated from different types of perturbation targets is a useful marker to estimate the degree of individual regions' and/or connections' vulnerability in the brain connectome.

Quantification of perturbation patterns provided by dynamical systems modeling of the brain may become a helpful tool when designing goal-directed interventions such as presenting sensory stimuli and interventions like applying transcranial magnetic stimulation. Furthermore, given the promising results of new closed loop deep brain stimulations (DBS) which are based on ongoing brain activity (Weerasinghe et al., 2018), we believe that a mathematical model of the brain oscillations will be helpful in designing an optimal stimulation strategy that provides detailed information about the particular state of the system which is requited. It may also allow us to better understand why some brain lesions cause cognitive and physical impairment that may become more severe over time while other lesion patterns have a much a better long-term outcome.

## 2. Materials and Methods

### 2.1. Data used and preprocessing

The primary data source for this study was the Human Connectome 500 subject release (Smith et al., 2013; Van Essen et al., 2012). Subject recruitment procedures and informed consent forms were approved by the Washington University institutional review board. The dataset is publically shared on the ConnectomeDB database (https://db.humanconnectome.org). Resting-state fMRI data were collected in two sessions, each session including two runs with phase encoding in either left-to-right or right-to-left directions. For our analysis, we used a single resting-state run, collected during 14.4 minutes with temporal resolution of 0.72 s. The entire dataset consisted of 1200 image volumes.

The dataset had been minimally-preprocessed (Glasser et al., 2013; Smith et al., 2013; Van Essen et al., 2012), which starts with gradient distortion correction and proceeds by realignment, bias field correction, spatial distortion removal, registration to standard Montreal Neurological Institute (MNI) space and intensity normalization (Glasser et al., 2013). Also, ICA+FIX ("FMRIB's ICA-based X-noiseifier") pipeline had been applied in order to automatically remove nuisance components (e.g., motion effects, non-neuronal physiology and scanner artefacts) from the fMRI data (Griffanti et al., 2014; Salimi-Khorshidi et al., 2014). Further preprocessing steps were added to the pipeline. The volumes collected during the first 10 seconds of the scan as well as the outlier volumes were discarded. Using the 3dDespike function in AFNI, outlier volumes were detected and interpolated from neighboring volumes. Next, the nuisance regression was performed using the global signal, mean white matter and cerebrospinal fluid (CSF) signals, as well as the 24 motion time-series (C.-G. Yan, Craddock, Zuo, Zang, & Milham, 2013) simultaneously with linear and quadratic detrending. In addition, the data was band-pass filtered (0.02–0.12 Hz) (Fukushima, Betzel, He, de Reus, et al., 2018). Moreover, the Freesurfer software (https://surfer.nmr.mgh.harvard.edu) was applied to parcellate the cortical surface of T1-weighted images into 68 anatomically segregated gyral-based regions-of-interest (Desikan et al., 2006). This data-driven parcellation is also known as the 'Desikan-Killiany' cortical atlas.

High-quality diffusion-weighted MRI data for the same 500 subjects from the HCP consortium (Glasser et al., 2013; Van Essen et al., 2012) was used for a streamline tractography on 68 cortical regions (Yeh, Wedeen, & Tseng, 2010). Next, we created subject-level weighted structural connectomes using measures of streamline density, computed by dividing the number of streamlines connecting two regions by the average of the volumes of the two interconnected regions to obtain streamline density (Hagmann et al., 2008; van den Heuvel, Kahn, Goñi, & Sporns, 2012; van den Heuvel & Sporns, 2011). For details on the processing steps of diffusion-MRI derived connectivity data we refer the reader to the earlier work (van den Heuvel & Sporns, 2011). We constructed a group-representative structural connectome by averaging the subject-level structural connectivity entries which had nonzero values for at least 60% of the subjects (de Reus & van den Heuvel, 2013), followed by resampling the data to follow a Gaussian distribution with $\mu = 0.5$ and $\sigma = 0.15$ (C J Honey et al., 2009; van den Heuvel et al., 2015).

### 2.2. Computational modeling of brain dynamics

In our system-level model of the brain, each region was modeled by a local mean-field model and they interact with each other through the structural connectome as previously described (Breakspear, 2017; Cabral et al., 2014; Deco & Jirsa, 2012; Deco et al., 2009; Gollo et al., 2015; C J Honey et al., 2009; Christopher J Honey et al., 2007). Accordingly, we model the local dynamics of each brain area with a modified Stuart-Landau equation. The Stuart-Landau equation describes the behavior of a nonlinear oscillating system near the Hopf bifurcation and it can be thought of as the principal model for nonlinear oscillators since it is the simplest possible model to describe amplitude dynamics (Röhm, Lüdge, & Schneider, 2018). When coupled together, the collective behavior of interacting oscillator systems has been shown to reproduce features of brain dynamics (Deco, Kringelbach, Jirsa, & Ritter, 2017; Freyer et al., 2011). Graph theory allows representing the interactions within the resulting complex network through a set of nodes which are connected by edges (Newman, 2003; Rubinov & Sporns, 2010; S H Strogatz, 2001). Since here we are modeling the neural activity of each brain region with the oscillations produced by an oscillator's model, the words "brain region", "node'' and "oscillator" will be used interchangeably in the text.

### 2.2.1. Local mean-field model

The Stuart-Landau equation describes the dynamic behavior of each oscillator j by:

$$\dot{z}_j = \left(a + i\omega_j - |z_j|^2\right) z_j \qquad (1)$$

where $z = r\, e^{i\theta} = r\cos\theta + ir\sin\theta$ is a complex number describing the state of the oscillator, $\omega \in \mathbb{R}$ is the frequency of each oscillator and the bifurcation parameter $a \in \mathbb{R}$ determines whether the oscillator is characterized by noisy fluctuations or exhibits oscillatory behavior.

By applying a slight change to the control parameter (a) of the nonlinear system, an abrupt qualitative change in the behavior of the system may occur, which is called bifurcation. For example, if the control parameter (a) in equation (1) moves from the negative to the positive domain, a critical point is reached when the pair of complex eigenvalues for the oscillator crosses the imaginary axis. At this bifurcation point, the equilibrium of the system loses its stability and a closed orbit (known as limit cycle) of radius $\sqrt{a}$ with a constant angular frequency ω develops, which is called a Hopf bifurcation (Supplementary Figure 1). The closed orbit in the phase space represents periodic behavior of the system (Hilborn, 2000; Kuramoto, 1984; Steven H. Strogatz, 2018).

Nonlinear dynamic models with stable limit cycles describe systems with self-sustained oscillations (i.e., oscillating behavior persists even in the absence of external periodic forcing or facing with slight perturbations). The non-linear system described above is easier to analyze if we rewrite the equation (1) in polar coordinates, which in turn give us the equations below:

$$\xRightarrow{z=re^{i\theta}} \begin{cases} \dot{r}_j = ar_j - r_j^3 \\ \dot{\theta}_j = \omega^0{}_j \end{cases} \quad (2)$$

If we translate the model stated in equation (2) to functional neuroimaging data, we can interpret the real part of the variable $z$ (i.e., $r\cos\theta$) as an indirect measure of brain activity acquired by the MR scanner, whereas the imaginary part serves as the hidden state of the system that is unobservable.

### 2.2.2. Modeling whole brain dynamics — Collective neurodynamics

The next step is to embed the local dynamics for each node as postulated in equations (1) and (2) into a large-scale model that encapsulates all nodes in the brain connectome. Then, the dynamics of whole brain is described by a system of coupled differential equations as given below:

$$\dot{z}_j = \left(a + i\omega_j - |z_j|^2\right)z_j + G\sum C_{ij}(z_i - z_j) + \beta\eta_j \quad (3)$$

In equation (3), $G \in \mathbb{R}$ is the global coupling parameter where the strength of coupling between regions is set by the structural connectivity matrix C (Arenas, Díaz-Guilera, Kurths, Moreno, & Zhou, 2008; Deco, Kringelbach, et al., 2017; Rodrigues, Peron, Ji, & Kurths, 2016). The constant G serves as a common tuning parameter that scales all the connection weights similarly.

In addition, additive Gaussian noise (denoted with $\eta_j$ with standard deviation of $\beta=0.02$ and implemented as Wiener process) was added to the differential equation of each oscillator to simulate the effect of random processes that occur in brain (e.g., stochastic effects of ion channels and heat), as well as inputs from sensory systems that have not been explicitly modeled (Roberts, Friston, & Breakspear, 2017).

### 2.2.3. Introducing frequency dynamics (freely running frequency evolution)

We describe the dynamics of angular frequency for every area *j* as:

$$\dot{\omega}_j = \omega^0{}_j - \lambda\, \omega_j \qquad (4)$$

where we introduce the coefficient $\lambda \in \mathbb{R}$ as the frequency lethargy. For brain region *j*, the angular frequency $\omega_j$ represents the free running frequency, and $\omega^0{}_j$ is its intrinsic frequency. The intrinsic frequency for brain region *j* was estimated from the actual BOLD signals of that particular region, as given by the median (across subjects) peak frequency of the mean (across voxels) BOLD signal.

Notably, in this study we introduce the possibility of frequency modulation of each oscillator, where the oscillating frequency can be modulated by its neighbors (we will later show that the frequency change for each region can be modulated by the net phase of all oscillators in its direct neighborhood). Thus, by combining the previous model of whole brain dynamics (as formulated in equation 3) with the suggested frequency modulation equation we arrive at the following coupled differential equations to describe the whole brain dynamics in the connectome:

$$\begin{cases} \dot{z}_j = \left(a + i\omega_j - |z_j|^2\right) z_j + G \sum C_{ij}(z_i - z_j) + \beta \eta_j \\ \dot{\omega}_j = \omega^0{}_j - \lambda\, \omega_j + m\, \psi_j \end{cases} \qquad (5)$$

The global phase of the neighbor ensemble for oscillator $j$ is denoted by phase angle $\psi_j$, which represents the interacting phases of the ensemble of all oscillators connected to this particular oscillator. Additionally, the model stated in equations (5) includes the frequency modulation coefficient $m$, which needs to be optimized for the BOLD data. The frequency modulation term was introduced to test the hypothesis that it could offer a suitable platform to understand the dynamic organization of synchronization patterns in the brain connectome.

In our model, frequency modulation is achieved through the structural connectivity matrix so that for each oscillator it limits the inclusion of oscillators in the neighbor ensemble to the ones which are directly structurally connected to that particular oscillator ($\psi_j = \sum C_{ij}\, \theta_i$). The phase of each oscillator (denoted by $\theta_i$) was calculated as the inverse tangent ($[-\pi/2, \pi/2]$) of the quotient of dividing the imaginary part by the real part of the system variable ($z_i$). Schematic overview of whole-brain modeling is illustrated in Figure 1.

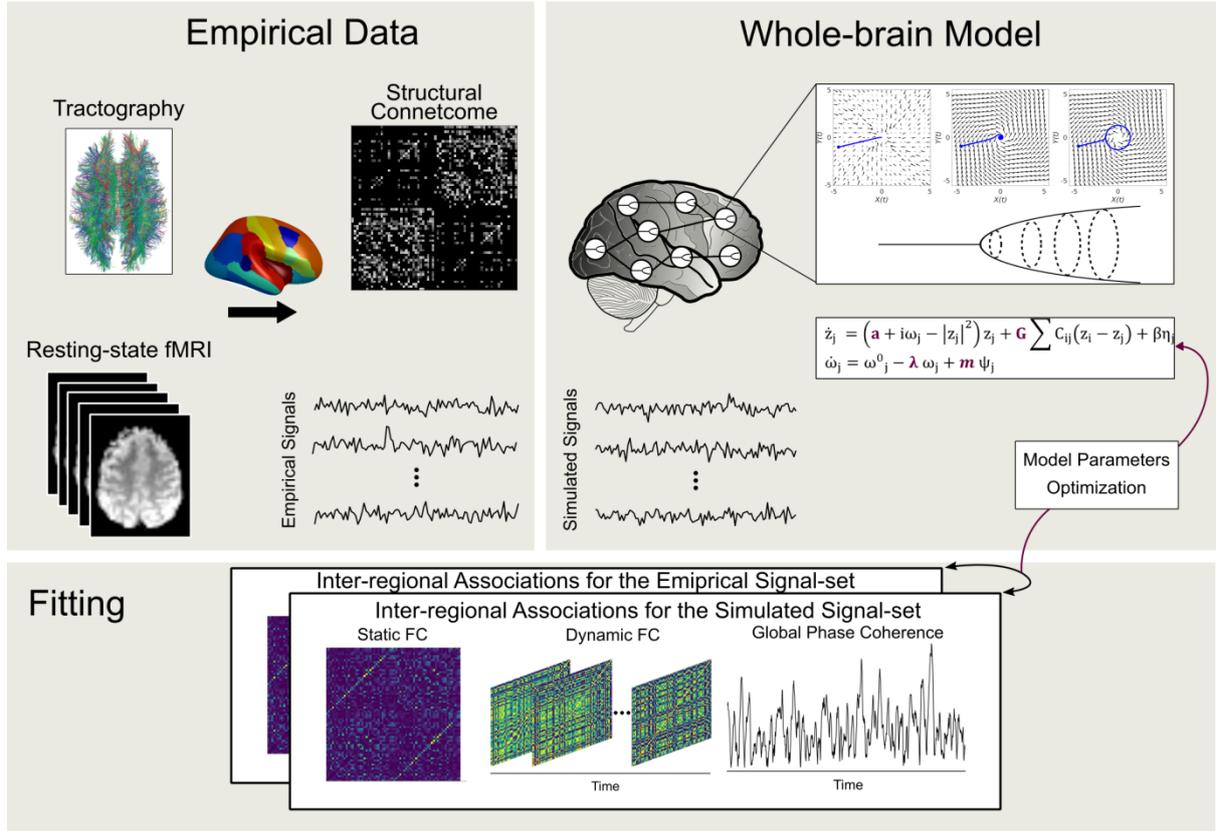

*Figure 1.* Construction of whole-brain network model. Using diffusion-MRI, streamline tractography was performed on 68 cortical regions. Subsequently, a group-representative structural connectome was built to govern the interaction strength between different regions. Resting-state brain dynamics was simulated using our adaptive frequency-based model. The model parameter set was tuned using the empirical data, based on several measures of functional connectivity and instantaneous coherence.

**2.3. Identification of the optimal working point for the model**

The global coupling parameter $G$ and the global bifurcation parameter $a$, frequency lethargy coefficient $\lambda$ and the frequency modulation coefficient $m$ are the control parameters of our model which are required to be optimized in order to find a working point where the simulated signals maximally fit the empirical BOLD signals. We performed a grid-search framework to estimate optimal values for model parameters. Maximum fitness was achieved by minimizing the dissimilarity between model-driven measures and the metrics extracted based on the empirical BOLD signals. Measures of interest included time-averaged measures like the Pearson correlation of static FC patterns and modularity which quantifies the level of segregation for the resting-state subnetworks, as well as three measures calculated based on instantaneous phase of signals. We computed a composite score by converting each metric to a unity-based normalized distance between model and empirical data and subsequently averaging them (Supplementary Equations). Applied methods for computation of these measures are summarized below:

### 2.3.1. Measures based on time-averaged functional connectivity

**Static functional connectivity patterns**

The matrix of correlation coefficients was computed from the pre-processed BOLD signals originating from all 68 regions-of-interest as included in the 'Desikan-Killiany' cortical atlas. A group-level FC matrix was calculated by averaging the z-transformed subject-level connectivity matrices. The Pearson correlation coefficient between the group-level FC matrix and the FC matrix obtained for the simulated signals was used as one of the measures indicative of similarity between empirical and simulated BOLD signals.

**Whole brain modularity**

Previous research have found functional networks/systems as ensembles of brain regions that co-activate during resting-state as well as during tasks (Smith et al., 2009). In order to associate every region of brain with a functional network, we first overlaid the 68 regions with the functional networks (so called "resting-state networks") which were previously defined based on the similarity of intrinsic FC profiles in 1,000 subjects (Yeo et al., 2011). Based on the percentage of overlap, we related each region with one of the functional networks. Accordingly, the 6 functional networks considered here included the default mode network (DM), the limbic network (LIM), the dorsal attention or control network (dTT/CONT), the salience or ventral attention network (SAL/vATT), the somatomotor network (SOM) and the visual network (VIS). Supplementary Figure 2 displays the layout of these functional networks. The used abbreviations for brain regions are shown in Supplementary Table 1.

Indeed, the whole-brain network can be divided into several modules (Betzel et al., 2014; Rubinov & Sporns, 2011) that are in good agreement with known functional systems (Meunier, Lambiotte, Fornito, Ersche, & Bullmore, 2009; Power et al., 2011). The presence of these modules that correspond to the functional systems is an indication that the simulated network adequately models some characteristics of brain organization.

Modularity quantifies the degree to which a network can be divided into the groups of nodes (i.e., modules) with stronger intra-module connections and weaker inter-module connections. For a FC matrix with positive weights, modularity is defined as:

$$Q = \frac{1}{v} \sum_{i,j} (w_{ij} - \frac{s_i s_j}{v}) \delta_{ij} \qquad (6)$$

where the $i$, $j$-th element of the FC matrix is denoted by $w_{ij}$, and $s_i = \sum_j w_{ij}$ is the nodal strength. The variable $v = \sum_{ij} w_{ij}$ is the overall weight of the network. The Kronecker delta function $\delta_{ij}$ is equal to one if the $i$-th and $j$-th nodes belong to the same module, and 0 otherwise. Hence, we computed the modularity ($Q$) of whole-brain network, assuming the abovementioned functional networks (DM, LIM, dATT/CONT, SAL/vATT, SOM, and VIS) being the network's modules.

### 2.3.2. Instantaneous phase-based measures

Here, we applied the Hilbert transformation to the regional BOLD signals to derive the analytic representation of the real-valued BOLD signals. We calculated the instantaneous phase of the analytic signal by computing the four quadrant inverse tangent ($tan^{-1}$) of the quotient formed by dividing the imaginary part by the real part of the BOLD signal.

**Dynamic functional connectivity patterns**

Computing the instantaneous phase synchrony (phase coherence) as a measure of time-varying FC offers single time-point resolution and has gained considerable attention in the recent literature (Omidvarnia et al., 2016; Pedersen, Omidvarnia, Walz, Zalesky, & Jackson, 2017; Ponce-Alvarez et al., 2015). The instantaneous FC for each pair of regions was defined by cosine similarity of the phases obtained from associated regions' signals. Thus, it was computed as $1 - |\sin(\Delta\theta)|$ at each time-point, where $\Delta\theta$ represents the instantaneous phase difference between two BOLD signals.

Next, the similarity between instantaneous FC measures of different time-points was calculated based on the cosine similarity between vectors created by applying half-vectorization on every instantaneous FC matrix. Next, we selected the cumulative distribution of the concatenated cosine similarities across all subjects as a measure of dynamic connectivity (Cabral, Kringelbach, & Deco, 2017; Deco, Kringelbach, et al., 2017; Deco & Kringelbach, 2016; Senden, Reuter, van den Heuvel, Goebel, & Deco, 2017). The same procedure was applied to the simulated data.

As a final step, we applied the Kolmogorov-Smirnoff test to evaluate the degree of agreement between dynamic FC patterns obtained by the empirical BOLD data and the results generated by our model. This comparison was performed for all tested combinations of the control parameters in the model.

**Macroscopic coherence of the model system**

At the edge of the critical point of the bifurcation in a system of coupled oscillators, there is a transition of the global attractor from incoherence to synchrony (Skardal, Ott, & Restrepo, 2011), which can be defined through the emergence of a macroscopic mean-field. This mean-field is computed as the centroid vector of the phase distribution as:

$$R = r\, exp(i\phi) = \frac{1}{N}\sum_{j=1}^{N} \exp(i\theta_j) \qquad (7)$$

The amplitude of the centroid vector (indicated by the scalar r) represents the phase divergence or uniformity of N oscillators and ϕ is the representative phase of the set of oscillators (Breakspear et al., 2010). Importantly, r describes the global phase coherence of the system at each time-point as it disappears when the phases of oscillators have large circular variance and approaches one when all the oscillators moving nearly in phase (Breakspear et al., 2010). It is customary to describe the global dynamic behavior of the ensemble using the mean and the standard deviation of r across time-points, which are referred to as the global synchrony and global metastability, respectively (Cabral, Hugues, Sporns, & Deco, 2011; Váša et al., 2015). Metastability refers to the existence of a form of "winnerless competition" between two apparently opposing tendencies, namely, a tendency of individual oscillators to couple with each other and coordinate globally for multiple functions, and a tendency to be independent to express their specialized functions (Kelso, 2008; Roberts et al., 2018; Tognoli & Kelso, 2014). We computed global synchrony and global metastability as indicative of macroscopic coherence of the whole-brain network model, for all tested combinations of the control parameters. In addition, we calculated these measures for the empirical BOLD signals (Supplementary Figure 3). We calculated the distance between empirical and simulated global synchrony and metastability in order to evaluate their agreement.

**2.4. Perturbation assessment**

We simulated perturbation to every individual brain region in our model by shifting the dynamic regime of the targeted region from the oscillatory dynamics domain (characterized by the estimated bifurcation parameter) to the noise-driven fluctuations. With regard to investigating the effects from perturbations, we took advantage of the mathematical representation of the brain network as a graph with a set of nodes symbolizing brain regions and edges denoting the mutual interactions among nodes (Newman, 2003; Rubinov & Sporns, 2010; S H Strogatz, 2001).

**2.4.1. Robustness — network breakdown under random failures versus targeted attacks**

Robustness refers to the capacity of a system to absorb disturbances caused by either internal or external faults and still maintain its basic structure and function, even if some nodes and edges may be missing (Barabási, 2016). It has been shown that failure of hub regions in the brain ("targeted attacks") have more detrimental effects on the network structure compared to "random failures" (Barabási, 2016). To test the performance of our model with regard to targeted attacks versus random failures, we simulated perturbation of every individual region by setting the bifurcation parameter to $a = -2$. An increasing fraction of regions, as denoted by $f$, were selected and perturbed, followed by measuring the size of largest strongly connected component (so called "giant component") formed in the network (Barabási, 2016). In an undirected graph, two nodes belong to the same component if there is at least one sequence of edges connecting them. The presence and absence of edges are denoted by binary edges. Therefore, giant component size was measured after applying a binary classification of the edges into two groups (0: disconnected or 1: connected) on the basis of a classification rule (i.e., threshold). We tried 100 different thresholds (0-1) for the binary classification of the static FC matrices which were computed based on either the empirical BOLD data (empirical FC) or the simulated BOLD data (using optimal parameter set).

Next, we computed the accuracy of the binary classification test at every threshold (Supplementary Figure 4). Accuracy was determined by dividing the number of correct assessments (number of true positives + number of true negatives) into the number of all assessments (number of true/false positives + number of true/false negatives). Correct assessments refer to the functional connections which are classified as connected edges in both empirical and simulated FC matrices. As illustrated in Supplementary Figure 4, the application of conservative thresholds resulted in that the accuracy of binarization increases at the expense of precision. Precision is the number of true positive assessments divided by the number of all positive assessments (number of true/false positives) returned by the classifier. Finally, after we fitted the cubic polynomial curve to the precision values (illustrated as the dotted green curve in Supplementary Figure 4), we estimated the location of the knee of both curves to be 0.27. Additionally, the intersection point of the two curves gives the threshold of 0.08 which is a more liberal threshold compared with the aforementioned knee point. It is worth mentioning that applying a common absolute threshold to the perturbation maps (as models of impaired brain connectome) is preferred to relative thresholding in the literature (Fornito, Zalesky, & Bullmore, 2016).

Testing the robustness of our modelled brain network was performed using two different perturbation strategies: (1) Simulating random failures by applying perturbations to an increasing fraction of regions which were randomly selected. This procedure was repeated many times (n = 2000) while the random number generator was reseeded at each iteration. (2) Simulating targeted attacks by applying perturbations to an increasing fraction of regions which were already sorted according to their degree of centrality in the structural connectivity matrix. That is, regions with a higher centrality were targeted for perturbation before the remaining regions. The degree of centrality of each region was measured as a composite hub-score that was calculated by averaging the unity-based normalized measures of nodal strength, betweenness centrality and closeness centrality (Freeman, 1977, 1978; Kaboodvand, Bäckman, Nyberg, & Salami, 2018; Rubinov & Sporns, 2010; Sporns, Honey, & Kötter, 2007; van den Heuvel & Sporns, 2013).

### 2.4.2. Vulnerability mapping — assessment of the functional connectivity changes subject to distributed failures

The susceptibility of a networked system to undergo significant changes in its function when confronted with different forms of disruption is called vulnerability. Applying perturbations to different regions of the network, and subsequently analyzing the perturbation patterns, is called vulnerability mapping which aims to locate weaknesses within the system (Gollo et al., 2018). After applying the perturbation to every individual region of the brain network separately, we ended up with 68 sets of whole-brain network time-courses, each set simulating the brain with malfunction in one of the 68 brain regions. In addition, we had one set of simulated signals corresponding with the healthy brain which was modeled at optimal working point (see above). First, we computed the static FC matrices for every set of whole-brain signals (68 sets corresponding to simulated perturbations and one set for the optimal working point). Subsequently, we computed the nodal strength, followed by measuring the relative difference of nodal strength vector derived for every perturbation set, from the nodal strength vector yielded by the unperturbed set. Hence, we obtained 68 relative difference vectors, each representing the percentage change in nodal strength patterns caused by an induced single-node perturbation. Then, we aggregated the positive and negative values of percentage changes, separately. This was done in order to measure two different types of nodal vulnerability: nodal hyper-connectivity (bias to increase FC) and nodal hypo-connectivity (bias to decrease FC). We computed the average of positive and negative entities, representing measures of nodal hyper-connectivity and hypo-connectivity.

In addition, inter-network FC between each pair of functional networks (DMN, LIM, CONT/dATT, SAL/vATT, SOM and VIS networks) was computed by averaging FCs among all node-pairs belonging to different functional networks (Kaboodvand et al., 2018). To compute the link vulnerability, we first subtracted the inter-network FC matrices derived for every perturbation set from the inter-network FC matrix yielded by the unperturbed set, followed by dividing the subtraction result into the FC matrix of the unperturbed set. Hence, we obtained 68 normalized divergence maps, each representing the percentage change in inter-network FC patterns caused by an induced single-node perturbation. Next, we aggregated the positive and negative values of normalized divergence maps, separately. This was done in order to measure two different types of link vulnerability: link hyper-connectivity and link hypo-connectivity risks. Therefore, we obtained measures of link hypo-/hyper-connectivity by averaging positive and negative entities independently.

### 2.4.3. Hazard mapping — assessment of the hazardousness of different brain regions

In addition to the three instantaneous phase-based measures which were used for finding the optimal model parameters, we calculated static FC-based measures including the global efficiency of whole brain network, system-wise local efficiency and the level of segregation for every functional system separately. Measures of the global and local efficiency were normalized by a matched (preserved degree distribution) random null model. Aforementioned measures were computed for the model of healthy brain (one set of whole-brain time-courses corresponding to optimal simulated brain) as well as for 68 models of the impaired brain (68 sets of whole brain time-courses corresponding to simulated local perturbations). Each measure was unity normalized across 69 observations. Subsequently, static FC-based measures were recruited to create a 13-dimensional feature vector for every simulated perturbed/unperturbed set of whole-brain signals. The vector space associated with these 69 feature vectors is called the feature space. In a similar way, we created a 3-dimensional feature space for instantaneous phase-based measures.

If we apply a perturbation to any region of the brain, it is likely to cause a feature space divergence from the optimal simulated brain network. We refer to the relationship between the location of each perturbation and the level of distance in feature space as hazard mapping, while the measured distance indicates the degree of hazardousness for that particular location. We computed the pairwise Euclidean distance between the feature vectors obtained from the simulated perturbation sets and the feature vector of simulated healthy brain to investigate how every region's failure contributes to increasing the dissimilarity between the simulated impaired brain and the simulated healthy brain. The level of distance was separately computed for the static FC-based measures and instantaneous phase-based measures. In the case of dynamic connectivity pattern, the Kolmogorov-Smirnoff distance was used as the numerical difference of their values. Afterwards, the two Euclidean distance measures were unity-based normalized and then summed together to create one distance measure for every perturbation.

A region shows a significant level of hazardousness if applying the perturbation to that particular region causes a substantial level of dissimilarity between the measures obtained from the simulated perturbation set and with the measures calculated for the simulated healthy brain. A brain region with the highest level of hazardousness may be interpreted as the region with the lowest level of fault tolerance.

## 3. Results

### 3.1. Model fitting and frequency dynamics

The control parameters of the model (i.e., $a, G, \lambda$ and $m$) were estimated in a grid-search framework, extended to a four-dimensional space. We performed a grid search in two steps using a different granularity in each step. First, we used a coarse-grained search that spanned a wide range of values for each parameter. Then, we used the results from the first search to perform a second, more fine-grained search for the optimal choices of $a, G, \lambda$ and $m$. From the first search, we found evidence supporting our hypothesis that resting brain operates at the edge of a critical point of bifurcation, with bifurcation parameters being close to zero produced a better fit of the model. In the fine-grained grid search, frequency modulation coefficient $m$ ranged from 0.1 to 0.2 in 11 steps and frequency lethargy $\lambda$ ranged from 0.2 to 1 with a step-size of 0.2. The grid-search for the global coupling $G$ included 15 evenly distributed values in the interval of 0.002 to 0.03. The spanning range for the global bifurcation $a$ included 71 values in the range from minus to plus 0.07.

Figure 2 illustrates the grid-search landscapes of global bifurcation $a$ and global coupling $G$ for five different measures of interest, separately for our proposed frequency modulation-based model versus coupled Stuart-Landau oscillators. The optimal choice of global coupling ($G$), bifurcation parameter ($a$), frequency lethargy $\lambda$ and frequency modulation $m$ is show as a white asterisk in the first panel of Figure 2 ($G = 0.01$, $a = 0.038$, $\lambda = 0.4$ and $m = 0.14$). The optimal parameter set is where the level of modularity of the model and the Pearson correlation between static FC derived from the model and empirical data is high. At the same time, the Kolmogorov-Smirnoff distance between the similarities of coherence measures, obtained from empirical BOLD data and simulated signals, as well as the differences of metastability and synchronization of simulated signals from the average metastability and synchronization measures of empirical data are considerably low. In order to provide evidence that increased number of model parameters is justified by goodness of fit, we computed the Akaike information criterion (AIC). Using the optimal parameter set, we re-computed the aforementioned composite similarity score, separately for each individual's empirical BOLD signals as a reference. The resultant distribution of composite scores was used to estimate the maximum value of the likelihood function for the model. Our results showed that the frequency modulation-based model had smaller absolute AIC value (-637.41) and performs better that the coupled Stuart-Landau oscillators (-889.28).

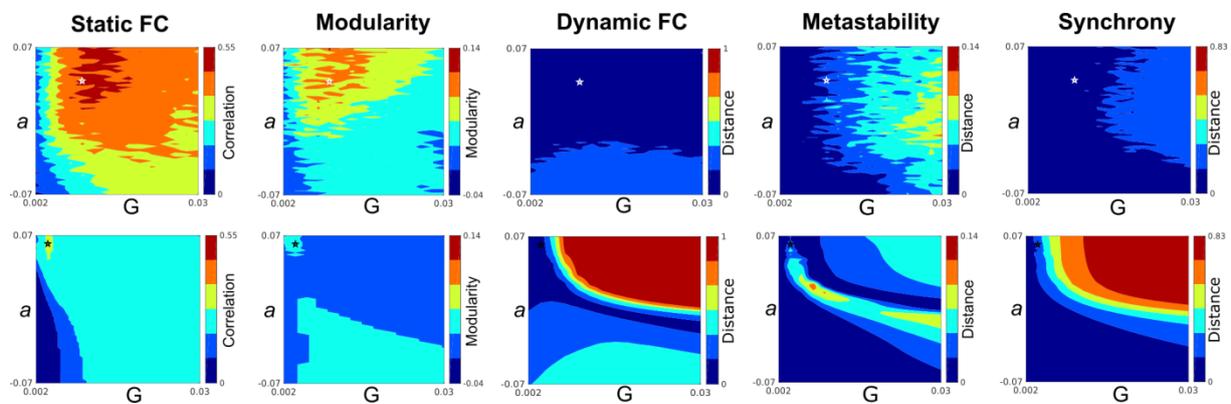

*Figure 2. Parameter space search. This figure shows the exploration of the parameter space defined by the bifurcation parameter **a** and global coupling **G**, separately for our suggested frequency modulation-based model (first row) versus coupled Stuart-Landau oscillators (second row). Measures in the upper panel are reported at the optimal values of frequency lethargy $\lambda = 0.4$ and frequency modulation $m = 0.14$. First column depicts the Pearson correlation between empirical and simulated static FC patterns for different pairings of bifurcation parameter **a** and global coupling **G**. Second column shows the whole brain modularity computed for the simulated static FC matrix. The Kolmogorov-Smirnoff distance between the similarities of coherence measures, obtained from the empirical BOLD data and simulated signals, as well as the differences of metastability and synchronization of simulated signals from the average metastability and synchronization measures of empirical data, are*

*respectively illustrated in columns 3-5. Measures associated with the optimal choice of global coupling (**G**), bifurcation parameter (**a**), frequency lethargy **λ** and frequency modulation **m** are shown as a white asterisk in the upper panel (**G** = **0.01**, **a** = **0.038**, **λ** = **0.4** and **m** = **0.14**), whereas the optimal point for the classic coupled Stuart-Landau oscillators is depicted by a black asterisk in the lower panel (**G** = **0.004**, **a** = **0.062**). See Supplementary Figure 5 for corresponding parameter space search including frequency lethargy **λ** and frequency modulation **m**.*

In the proposed model of the brain oscillations, the frequency of each oscillator is modulated by the net phase of the neighbor ensemble. Figure 3a shows the positive association between net phase of the neighbor ensemble and change of frequency in the case of right precuneus (chosen for illustration purposes). The frequency of an oscillator undergoes the highest change when net phase of the neighbor ensemble reaches its extremum. In other words, the highest frequency change rate is yielded for oscillator $j$ when the net phase of its neighbor ensemble $\psi_j = \sum C_{ij} \theta_i$ reaches its extremums. On the other hand, phase of each neighbor oscillator (denoted by $\theta_i$) is related to the activity of that particular oscillator (i.e., $r_i \cos\theta_i$). Figure 3b illustrates how phase of left posterior cingulate cortex (an example of a neighbor of the right precuneus cortex) is associated with its activity. The phase of oscillator begins to grow when the magnitude (i.e., absolute value) of its activity starts to shrink and it approaches the maximum value (illustrated by light green upward-pointing triangles in Figure 3b) when the magnitude of its activity is passing zero (illustrated by empty circles in Figure 3b). In other words, when the absolute value of overall activity in the neighborhood of oscillator $j$ starts to decline, the net phase of the neighbor ensemble ($\psi_j$) begins to grow until the overall neighborhood activity passes zero and starts to regrow again. Therefore, the net phase of the neighbor ensemble ($\psi_j$) reaches its maximum value at the point where the overall neighborhood activity is at the minimum level. In conclusion, the frequency of an oscillator undergoes the highest change when the net activity of its neighbor ensemble is in the vicinity of zero, so that the speed of frequency changes reaches its maximum positive value (speed-up) when the net neighborhood activity gets close to pass the zero line, whereas it gets its maximum negative value (slow-down) when the net neighborhood activity has just started to grow.

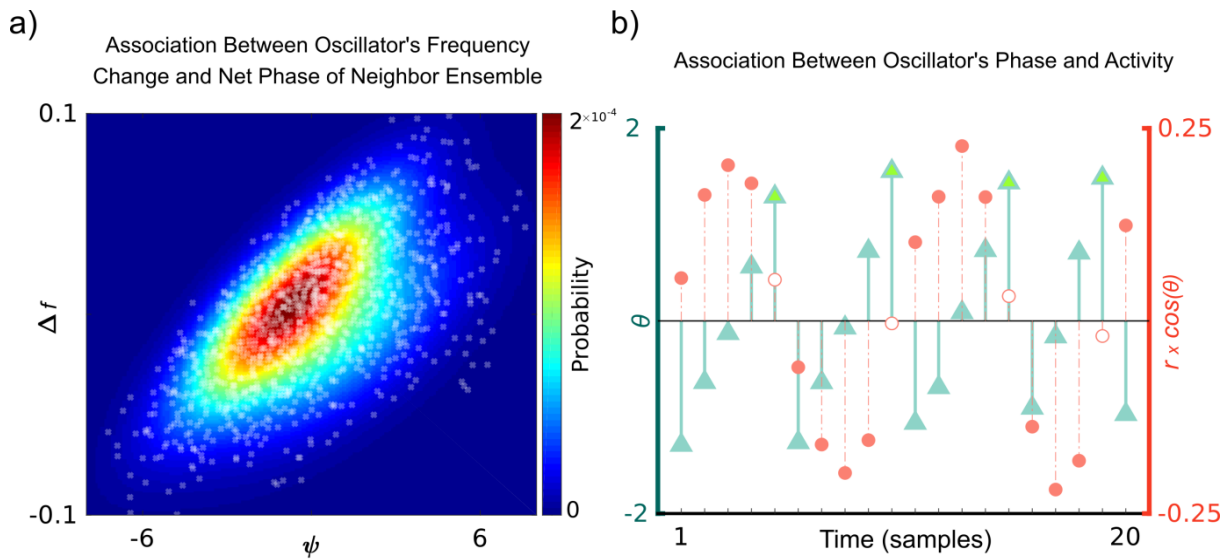

*Figure 3. Frequency dynamics in the proposed model of brain oscillations. (a) There is a positive association between the frequency change of an oscillator and net phase of its neighbor ensemble. The oscillator modeling the right precuneus cortex was selected for illustration purpose. The probability distribution for the Gaussian mixture model across two components (i.e., net phase of neighbor ensemble and change of frequency) is depicted in the first panel. (b) Association between the phase ($\theta$) of the oscillator modeling the left posterior cingulate cortex (an example neighbor of the right precuneus) and its level of activity ($r \times \cos\theta$). The phase of oscillator begins to grow when the magnitude of its activity starts to shrink and it approaches the maximum value (illustrated by light green upward-pointing triangles in Figure 3b) when the magnitude of its activity is going to pass the zero value (illustrated by empty circles in Figure 3b).*

## 3.2. Robustness, vulnerability and hazard mapping for the whole-brain network model

As a measure of network robustness, Figure 4 displays the fraction of brain regions that belong to the giant component after applying either random or targeted perturbations to an $f$ fraction of regions. The size of the giant component at every value of $f$ was divided by the actual size of giant component, which provides a relative measure of giant component size. From Figure 4 it can be seen that in the face of random failure, the fragmentation process is gradual. However, the whole-brain network has a lower tolerance when facing with selective attacks to hub regions. Thus, our simulated brain network shows lower degree of robustness for targeted attacks versus random failures.

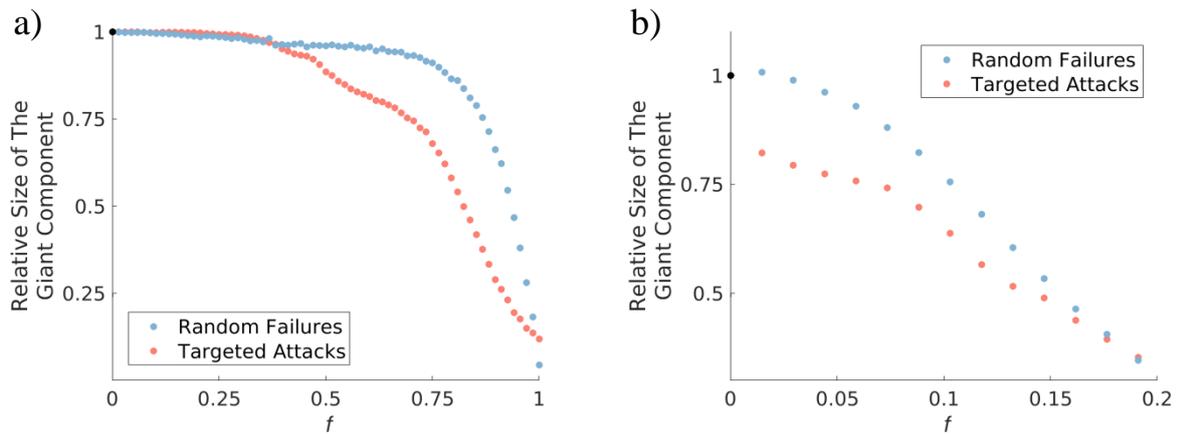

*Figure 4. Network breakdown under random failures versus targeted attacks is illustrated for both a liberal (absolute threshold = 0.08) and restrictive (absolute threshold = 0.27) thresholding strategy, respectively in panels (a) and (b). In the case of random failures, the fraction of nodes that belong to the giant component is computed after an f fraction of nodes are randomly selected for perturbation. For a targeted attack, we calculate the fraction of nodes that belong to the giant component after an f fraction of nodes are perturbed in a decreasing order of their hub-score, so that we start with the node with highest hub-score, followed by the next highest etc. The procedure of simulating random failures was repeated*

*2000 times which resulted in a smoothed plot. However, the Savitzky-Golay filter was applied to the values from targeted attacks, to enable the reader to easily see the general pattern of decline.*

Applying an *in silico* perturbation protocol to our proposed model enabled us to quantify network vulnerability in the form of hypo-/hyper-connectivity risk rate for either different brain regions (Figure 5; upper panel) or FCs between different functional systems (Figure 5; lower panel). Sub-regions of the cingulate cortex, the temporo-parietal junction (i.e., banks superior temporal sulcus), the parahippocampal cortex (encompassing the parahippocampal gyrus and the fusiform gyrus), the inferior temporal gyrus, the middle frontal gyrus, the inferior parietal cortex (including the inferior parietal gyrus and the angular gyrus) showed a strong risk for hyper-connectivity (Figure 5; uppermost panel, left column). Also, the posterior subdivision of inferior frontal gyrus (pars opercularis) in the right hemisphere, as well as the middle subdivision of inferior frontal gyrus (pars triangularis) in the left hemisphere had strong tendency to increase FC (Figure 5; uppermost panel, left column). On the other hand, we observed that the regions with the strongest hypo-connectivity risk were located in the posteromedial visual system, frontal pole, medial and lateral orbital frontal cortex, right supramarginal gyrus, right inferior frontal gyrus (pars triangularis and pars orbitalis) and right caudal middle frontal gyrus, and to a lesser extent in the right sensory-motor cortex (post central and precentral gyrus), the right insular cortex and the right superior temporal gyrus as well as left paracentral lobule (Figure 5; upper panel, right column). Furthermore, we observed the highest hyper-connectivity risk for the FC between the VIS system and the SOM and LIM systems, as well as for the CONT/dATT and LIM functional connectivity (Figure 5; lower panel, left column). Additionally, we observed considerable hyper-connectivity risk for the FC between attentional subsystems (SAL/vATT and CONT/dATT), as well as between the SAL/vATT system and the VIS and DM systems, and to a lesser extent between the LIM and SOM systems. The strongest hypo-connectivity risks were found between the DM and VIS, LIM and SAL/vATT, and between the CONT/dATT system with SOM and VIS systems (Figure 5; lower panel, right column).

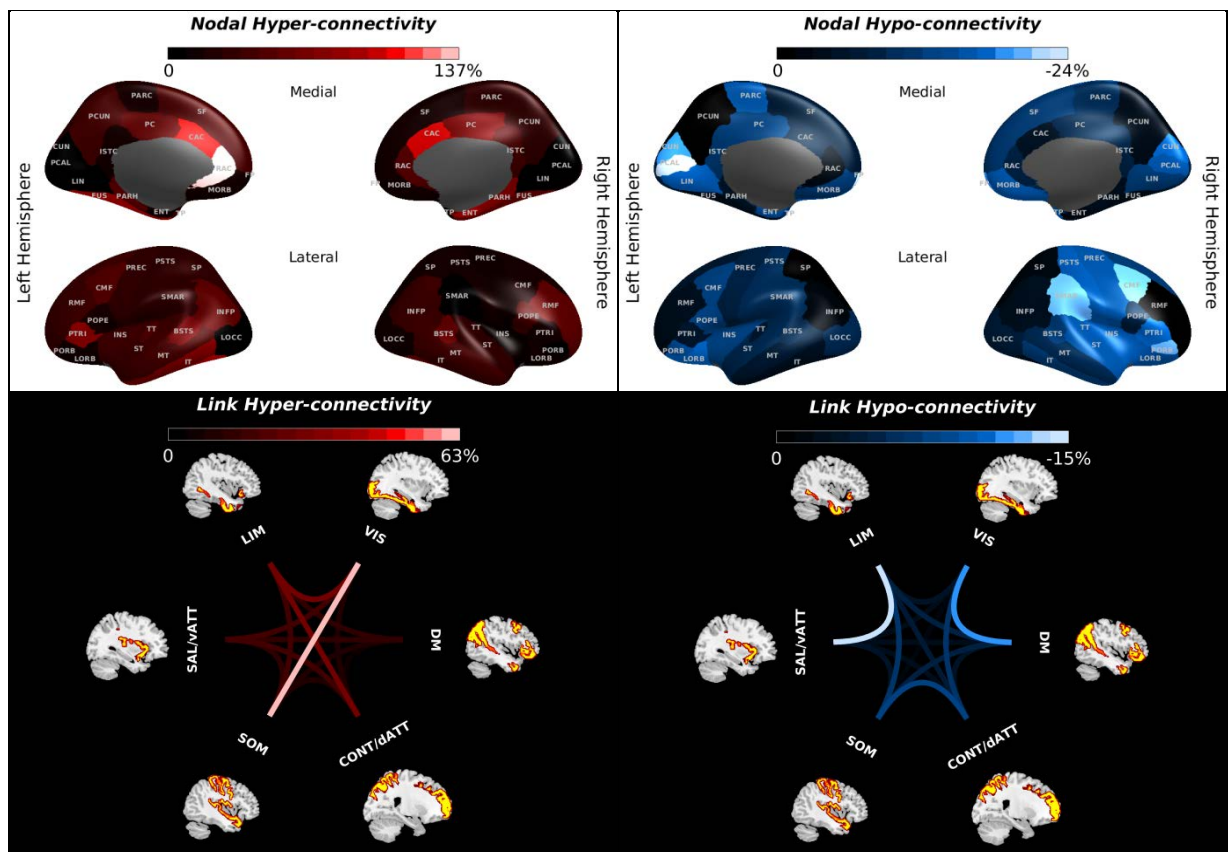

*Figure 5. Vulnerability mapping. The upper panel illustrates the vulnerability of different regions. The level of hyper-connectivity risk, which measures the tendency to increase FC in face of distributed failures in the whole brain network, is depicted in the left column, whereas the hypo-connectivity rate as indicative of the tendency to have decreased FC in face of distributed failures in the whole brain network is shown in the right column. The upper panel of this figure shows the nodal risk rate for hyper- and hypo-connectivity in color-format for every node listed in Supplementary Table 1. The lowermost panel of this figure illustrates the vulnerability of inter-network FCs by color-coded links between different resting-state networks. See Supplementary Table 1 for a list of the abbreviated node names. DM, default mode network (DM); LIM, the limbic system (LIM); dATT/CONT, the dorsal attention or control network; SAL/vATT, the salience or ventral attention network; SOM, the somatomotor network; VIS, the visual network.*

Finally, we examined the effects of malfunctions in different brain regions when a combination of local and global measures of brain network were taken into account (hazardousness mapping). The results of the hazardousness mapping are shown in Figure 6. In relating the hazardousness measures and the key nodal centrality measures, we found a significant correlation for the clustering coefficient (r = -0.33, *p* = 0.008), degree (r = 0.355, *p* = 0.006) and strength (r = 0.39, *p* = 0.004). Measure of association with hazardousness for the local efficiency was not significant (r = - 0.20, *p* = 0.094). Reported *p* values are FDR-corrected across four comparisons.

The analysis of the size of the damage inflicted on the network by applying perturbation to individual nodes showed that the superior parietal cortex (also known as dorsal attention system) and the left cuneus cortex, and to a lesser extent the precuneus cortex and the entorhinal cortex were the most critical regions for maintaining adequate network communication. In addition, we observed a high level of hazardousness for some regions in the right hemisphere such as the inferior parietal cortex (including the inferior parietal gyrus and the angular gyrus), the isthmus – cingulate cortex, the middle frontal gyrus, the insular cortex and the caudal anterior-cingulate cortex. Upon closer inspection, the medial orbital frontal cortex and the right posterior-cingulate cortex, as well as the left postcentral gyrus and right parahippocampal gyrus had considerable increased levels of hazardousness (Figure 6). Malfunction in any of the aforementioned areas resulted in a considerable divergence in the brain network characteristics.

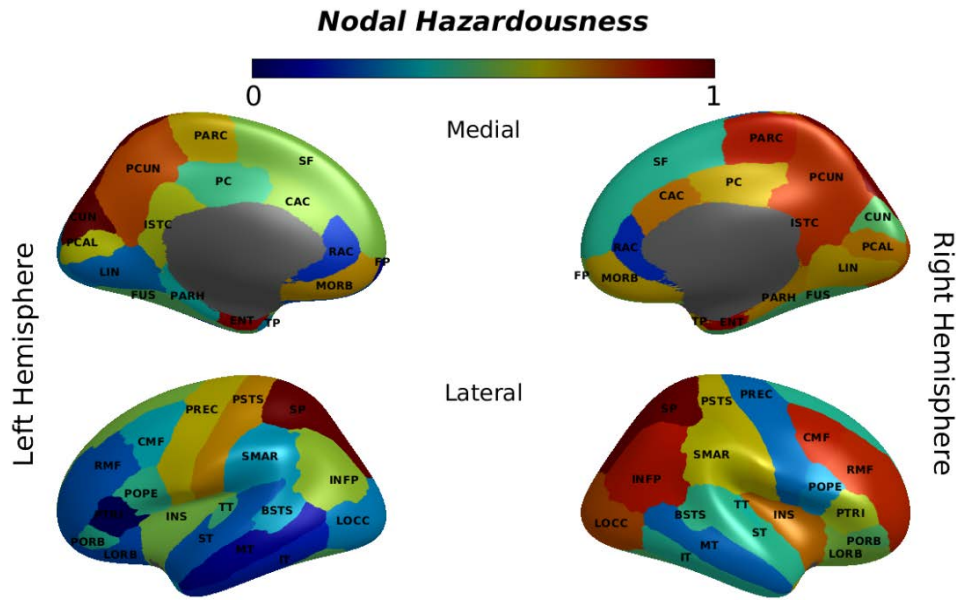

*Figure 6*. Hazardousness mapping. The degree of hazardousness for a region/node was measured as the degree of feature space divergence from the optimal simulated brain network caused by applying perturbation to that particular node. Normalized measures of distance in feature space are depicted in color-format for every node as listed in Supplementary Table 1. See Supplementary Table 1 for a list of the abbreviated node names.

## 4. Discussion

### 4.1. Frequency dynamics

Based on our constructed model, the oscillatory frequencies undergo the highest change when the activity in the neighborhood is in the vicinity of zero (Figure 3b). In other words, the slope of frequency change takes its maximum positive value (i.e., speeds up) when the input signals to that region are subsiding, while it experiences a slowdown when the input signals start to grow. Intuitively, this phenomenon may be likened to tuning a radio receiver, to enable it to receive incoming signals from the neighborhood. Supplementary Figure 6b illustrates the working frequency range of oscillators.

Of note, we know from the literature that the frequency spectrum of intrinsic oscillations of the brain has a time-varying nature (C. Chang & Glover, 2010; Yaesoubi et al., 2015), and notably we had hypothesized that the regional frequency is modulated by the activity of its neighbor regions. We were inspired by two famous electrophysiology theories: firstly, neurons acting as integrators, so that they sum or average their inputs to generate action potentials (Abeles, 1982; König, Engel, & Singer, 1996; Salinas & Sejnowski, 2001). Secondly, neurons being exquisitely sensitive to certain temporal input patterns, giving rise to oscillatory activity or switching from one oscillatory regime to another (Salinas & Sejnowski, 2001). Moreover, a previous science paper (Mukamel et al., 2005) has revealed that fMRI signals can provide a reliable measure of the firing rate of human cortical neurons.

Our model simulates ultra-slow oscillations. The power spectral density as well as the distribution of estimated intrinsic frequencies of these oscillations has been presented in Supplementary Figure 6a. Of note, the objective of our modeling study was mainly to reproduce the inter-relationships (i.e., FC and instantaneous coherence) of the human brain, not the frequency dynamics of the individual oscillators, notably only based on a fixed structural connectome (without modifying SC using the FC). To tackle this, we used a cost function that was based on inter-regional associations in the empirical data and not the intrinsic frequency, given a fixed SC matrix and a fixed level of activity for all brain regions (i.e., one common bifurcation parameter). Otherwise, tuning the local bifurcation parameters to reproduce the region-specific proportions of power in a narrow band (0.04–0.07 Hz; for example see, (Senden et al., 2017)) could increase the similarity between frequency characteristics of the simulated data and the empirical data, at the cost of introducing the possibility of overfitting due to optimization of several local bifurcation parameters for different brain regions. Furthermore, using one common bifurcation parameter for all different oscillators seemed the best fit for our later aim of applying *in silico* perturbations to the model.

### 4.2. Perturbation assessment

We simulated the effects of brain lesions by shifting the associated regions' dynamical regime to elicit noisy behavior, rather than removing nodes or links. A comparable protocol was previously used to temporarily either promote or disrupt synchronization in random brain regions and subsequently assess the recovery back to baseline (Deco et al., 2018). Of note, a previous study which optimized the region specific (also known as local) bifurcation parameters by fitting the normalized power for each intrinsic frequency band to the normalized power observed for empirical signal, could provide evidence that most brain regions in patients with Parkinson's disease had negative values of the bifurcation parameter representing neuronal noise corresponding to an asynchronous firing of neurons (Saenger et al., 2017).

#### 4.2.1. Robustness of the whole-brain network model

We observed that our model of brain has a lower tolerance when facing selective attacks to central regions. However, in the face of random failures, the fragmentation process was gradual. To the best of our knowledge, no previous study has evaluated robustness of brain (here measured as giant component size) by perturbing regions *in silico*. Removing nodes or links (as a way to simulate effects of structural lesions) is the most commonly used method for investigation of brain robustness. Since hub scores were calculated based on the actual structural connectome and perturbations were simulated by applying bifurcation-induced shifts in the dynamical regime, this finding may further validate the model and the recruited perturbation strategy presented here.

### 4.2.2. Hazardousness mapping

The hazard rates induced by primary failures of individual nodes provide relevant insights not only into the size of the damage inflicted on the network by individual nodes, but also into the potential origins of disease. Our results regarding nodal hazardousness rates are interesting in the context of the question why failures in some cases might lead to long-lasting impairment and disease, while other might not. To this end, we observed the highest level of hazardousness for the CONT/dATT system, which mainly covers the superior parietal cortex, the right caudal middle frontal gyrus and the bilateral rostral middle frontal gyrus, as well as for the posteromedial subsystem of the DM network (particularly the precuneus cortex, the inferior parietal cortex and the isthmus – cingulate cortex). In addition, the LIM system, in particular the entorhinal cortex was found to be a hazardous region. The SAL/vATT system (mostly including the medial subdivisions like posterior cingulate cortex and caudal anterior cingulate, but also the insular cortex) was also found to cause substantial levels of damage in case of malfunction (Figure 6). These results substantiate the important role of aforementioned regions for global coordination of information flow across the whole-brain network.

It is also of interest to relate our observation of a high level of hazardousness for the posteromedial subsystem of the DM network, to its important role in integrating the bottom-up attention with behaviorally related information from memory and perception (Andrews-Hanna, Smallwood, & Spreng, 2014). Indeed, previous studies have suggested that the ventral posteromedial cingulate cortex (referred to as the isthmus-cingulate cortex in the Desikan-Killiany atlas) has dense FC that essentially is restricted to the DM network (Kaboodvand et al., 2018; Leech, Kamourieh, Beckmann, & Sharp, 2011; Leech & Sharp, 2014; Spreng, Sepulcre, Turner, Stevens, & Schacter, 2013; Tomasi & Volkow, 2010). Hence, it may be said that it serves as the key provincial hub for the DM network, in that it mediates a large proportion of the traffic that facilitate interactions within this system (Kaboodvand et al., 2018). Moreover, the precuneus/posterior cingulate cortex has been identified as a critical connector hub (Achard, Salvador, Whitcher, Suckling, & Bullmore, 2006; Fransson & Marrelec, 2008; Leech et al., 2011; Spreng et al., 2013; van den Heuvel & Sporns, 2013; Zuo et al., 2012), with dense connections spreading across the whole brain network (Fransson & Marrelec, 2008; Hagmann et al., 2008). FC patterns of the posteromedial subsystem of the DMN have a well-established association with mild cognitive impairment and Alzheimer's disease (Bai et al., 2009; Buckner et al., 2005; Celone et al., 2006; Greicius, Srivastava, Reiss, & Menon, 2004). Positron emission tomography imaging in Alzheimer's disease have shown high amyloid-$\beta$ deposition in the prominent hubs located in posteromedial cingulate, lateral temporal, lateral parietal, and medial/lateral prefrontal cortices (Buckner et al., 2009; Hedden et al., 2009). There is some evidence suggesting that high level connectedness which is required for large information transfer in the hubs, may cause in augmentation of the underlying pathological cascade in AD (Buckner et al., 2009).

Given that our finding of regions with high hazardousness in the parietal association cortex was predominately right lateralized, it is worthwhile to inquire about the reason for this asymmetry. Indeed, previous literature suggests that lesions of the parietal association cortex in the right hemisphere are more detrimental whereas that left parietal lesions tend to be compensated by the intact right hemisphere (Purves et al., 2004). Notably, contralateral neglect syndrome is specially associated with the damage to the right parietal association cortex. There is evidence suggesting that the right parietal cortex mediates attention to both left and right halves of the body and the extra-personal space, whereas the left hemisphere mediates attention primarily to the right side (Purves et al., 2004). There is also strong evidence for the belief that the right middle fontal gyrus is a site of convergence of the dorsal and ventral attention systems, by acting as a circuit-breaker for interrupting ongoing endogenous attentional processes in the dorsal network and reorienting a person's attention to an exogenous task-relevant stimulus (Corbetta, Kincade, & Shulman, 2002; Fox, Corbetta, Snyder, Vincent, & Raichle, 2006; Japee, Holiday, Satyshur, Mukai, & Ungerleider, 2015). Therefore, the right middle fontal gyrus has control over both ventral attention and dorsal attention networks and would be responsible for the flexible modulation of internal and external attention (Corbetta et al., 2002; Fox et al., 2006; Japee et al., 2015).

We found high level of hazardousness for the entorhinal cortex. This finding resonates well with the previous studies that have shown that impairment of the entorhinal cortex is persistently reported in patients with Alzheimer's disease (Van Hoesen, Hyman, & Damasio, 1991), schizophrenia (Arnold, Ruscheinsky, & Han, 1997), as well as in cases of traumatic brain injury and stroke. Moreover, entorhinal damage is believed to cause interference with sensory integration and also cause memory deficits (in particular spatial learning impairment (Van Hoesen et al., 1991)). Moreover, it has been suggested that the age-related entorhinal cortex thinning occurs before hippocampal volume loss and FC decline in the DM system, which impacts communication between medial temporal lobe with the cortical DM system, contributing to age-related memory deficits (Tisserand, Visser, van Boxtel, & Jolles, 2000; Ward et al., 2015). Of note, it was recently discussed that the entorhinal cortex and retrosplenial cortex (referred as the isthmus – cingulate cortex in the Desikan-Killiany atlas) are sequential interfaces between the medial temporal lobe subsystem of the DM system and cortical regions of the DM system (Kaboodvand et al., 2018). In particular, it was shown that the entorhinal cortex mediates the interaction between medial temporal lobe and the retrosplenial cortex. Therefore, the dynamic coupling/decoupling of the medial temporal lobe from the cortical DM system (Huijbers, Pennartz, Cabeza, & Daselaar, 2011; Vannini et al., 2011; Young & McNaughton, 2009) is mediated by the entorhinal cortex and the retrosplenial cortex together (Kaboodvand et al., 2018). Thus, previous research on the important role for the entorhinal cortex in the brain's network is in well agreement with our findings of an increased level of hazardousness for this region.

Additionally, we found a considerable level of hazardousness for right anterior-cingulate cortex. The anterior-cingulate cortex is one of the key integrative brain hubs (Lavin et al., 2013), and it has been suggested to play important role in high-level cognitive processing, outcome monitoring, action planning and emotion processing (Fernández-Matarrubia et al., 2018; Lavin et al., 2013). Furthermore, it acts as a core component in fronto-striatal circuitry. There is evidence that abnormal functioning of right anterior cingulate may have a pathophysiologic role in speech impairment (C.-C. Chang, Lee, Lui, & Lai, 2007) and in the cognitive and emotional impairment related to attention deficit hyperactivity disorder (Tian et al., 2006), schizophrenia (H. Yan et al., 2012) and panic disorder (Shinoura et al., 2011). Notably, there is some evidence in support of FC changes between medial frontal and anterior cingulate in different dementia groups (Fernández-Matarrubia et al., 2018).

### 4.2.3. Vulnerability mapping

In response to distributed malfunctions (i.e., heterogeneous locations of injury), we observed that functional connections of our brain are obviously more susceptible to hyper-connectivity than hypo-connectivity, particularly in brain regions with high connectedness. Previous studies have repeatedly reported increased functional connectivity after traumatic brain injury as a compensatory brain response to make up for physiological disturbances (Bharath et al., 2015; Hillary et al., 2014; Iraji et al., 2016; Nakamura, Hillary, & Biswal, 2009), particularly so for regions with high connectedness (Hillary et al., 2014). Basically, the observation that "the rich get richer" is in line with the preferential attachment theory underlying scale-free network development, which suggests that new connections mostly occur in the regions with high connectedness (Barabási, 2016). Accumulating evidence have suggested that hyper-connectivity is a common response to neurological disruption in brain insults such as traumatic brain injury, mild cognitive impairment, Alzheimer's disease, multiple sclerosis and epilepsy (Hillary et al., 2014, 2015). Our nodal vulnerability analysis revealed that the most prominent cases for hyper-connectivity belong to the regions in the default mode, salience, and control systems. More specifically, the anterior-cingulate cortex, the inferior parietal cortex, the left precuneus cortex and also the posterior-cingulate cortex and the middle frontal gyrus, which is consistent with the previous literature on traumatic brain injury (Hillary et al., 2014; Iraji et al., 2016; Mayer, Mannell, Ling, Gasparovic, & Yeo, 2011). For example, increased connections in the acute phase of injury have been reported for the left cingulate gyrus, the left precuneus and the right prefrontal cortices (Bharath et al., 2015; Hillary et al., 2014).

Our results of an increased vulnerability in the fronto-parietal regions is perhaps not overly surprising given that fronto-parietal regions are involved in the top-down attentional control. Interestingly, hyper-connectivity of fronto-parietal regions in the traumatic brain injury patients is attributed to the increased awareness of the external environment. These observations may explain the reports of cognitive fatigue in these patients (Bharath et al., 2015; Muller & Virji-Babul, 2018; Shumskaya, Andriessen, Norris, & Vos, 2012).

Moreover, previous results suggests that the traumatic brain injury-related hyper-connectivity of the DM system is beneficial for cognitive function (Sharp et al., 2011), in that patients with stronger FC showed the least amount of cognitive impairment. A previous longitudinal study found that during the recovery phase after injury, the weight of network connections diminishes. This finding implies that over the course of recovery, the functional connectome of the injured brain begins to approximate the healthy functional connectome (Nakamura et al., 2009).

Furthermore, there are reports of the decreased FC in the literature which are predominantly right-sided, involving the right frontal and parietal regions (Bharath et al., 2015; Borich, Babul, Yuan, Boyd, & Virji-Babul, 2015; Mayer et al., 2011). For example, the hypo-connectivity for the right supramarginal gyrus and the right middle frontal gyrus, has been reported for the patients with mild brain injury (Borich et al., 2015; Mayer et al., 2011). The lack of dynamic flexibility after traumatic brain injury has been linked to the observed hypo-connectivity of right frontal eye field (Muller & Virji-Babul, 2018).

Our vulnerability mapping suggests that the SAL/vATT system, the CONT/dATT system and the DM system included the regions that had the highest hyper-connectivity risks (e.g., the posterior/anterior- cingulate cortices, the rostral middle frontal and the inferior parietal cortex). On the other hand, we found that the medial visual network, regions of the right parietal lobe (the right supramarginal gyrus and postcentral gyrus), as well as regions of the frontal lobe (caudal middle frontal gyrus, medial and lateral orbital frontal cortex, frontal pole, pars orbitalis subdivision of the inferior frontal gyrus) had noticeable risks of hypo-connectivity.

Our results regarding the vulnerability of inter-system connections are also in line with the published literature on the effects of injury to brain networks. For example, a previous study of mild traumatic brain injury has reported functional hyper-connectivity for the FC of posterior-cingulate cortex with the frontal eye fields, the dorsolateral prefrontal cortex, the associative visual cortex, the somatosensory association cortex and the premotor cortex, as well as for the FC of occipital lobe with the frontal lobe (Iraji et al., 2016). Our results also indicated considerable level of hyper-connectivity for the SAL/vATT system, encompassing the posterior-cingulate cortex. Of note, there is a body of evidence regarding the hyper-connectivity within the frontal lobe (Bharath et al., 2015; Borich et al., 2015; Hillary et al., 2014; Mayer et al., 2011).

These previous clinical observations give some support to our results of significant hyper-connectivity risks for the FC of SAL/vATT system (which encompasses the posterior-cingulate cortex) with the DM, CONT/dATT, SOM and VIS systems, as well as strong hyper-connectivity between the VIS and LIM systems. Moreover, previous reports of the hyper-connectivity within the frontal lobe (Bharath et al., 2015; Hillary et al., 2014) are of relevance to our observation of strong hyper-connectivity between the CONT system and the limbic or the SAL/vATT systems. Abnormal hyper-connectivity between the top-down attentional systems has been regarded as the underlying cause for some post-traumatic brain injury symptoms, such increased distractibility and cognitive fatigue (Borich et al., 2015; Mayer et al., 2011; Shumskaya et al., 2012).

Taken together, our findings regarding hyper/hypo-connectivity risks in response to distributed brain failures are largely in agreement with the observations reported in the aforementioned clinical studies. Thus there are good reasons to believe that the *in silico* perturbation assessment of the whole-brain dynamical connectome can provide valuable information in predicting re-organization of brain connectivity in response to neurological dysfunctions. In addition, we provided novel evidence for dissimilar susceptibility of FCs when facing heterogeneous failures in the brain network. Moreover, our proposed hazardousness map may serve as a useful index for predicting the impact of specific failures on network characteristics and a particular damage's contribution in potential long-lasting impairment and disease (Barabási, 2016).

**Acknowledgement**

P.F. was supported by the Swedish Research Council (grant No. 2016-03352) and the Swedish e-Science Research Center. Data were provided by the Human Connectome Project, WU-Minn Consortium (Principal Investigators: David Van Essen and Kamil Ugurbil;

1U54MH091657) funded by the 16 NIH Institutes and Centers that support the NIH Blueprint for Neuroscience Research; and by the McDonnell Center for Systems Neuroscience at Washington University. The funders had no role in study design, data collection and analysis, decision to publish, or preparation of the manuscript. The authors thank Behzad Iravani for his valuable methodological advice.

References

Abeles, M. (1982). Role of the cortical neuron: integrator or coincidence detector? *Israel Journal of Medical Sciences*, *18*(1), 83–92.

Achard, S., Salvador, R., Whitcher, B., Suckling, J., & Bullmore, E. (2006). A resilient, low-frequency, small-world human brain functional network with highly connected association cortical hubs. *The Journal of Neuroscience*, *26*(1), 63–72. https://doi.org/10.1523/JNEUROSCI.3874-05.2006

Aerts, H., Fias, W., Caeyenberghs, K., & Marinazzo, D. (2016). Brain networks under attack: robustness properties and the impact of lesions. *Brain: A Journal of Neurology*, *139*(Pt 12), 3063–3083. https://doi.org/10.1093/brain/aww194

Andrews-Hanna, J. R., Smallwood, J., & Spreng, R. N. (2014). The default network and self-generated thought: component processes, dynamic control, and clinical relevance. *Annals of the New York Academy of Sciences*, *1316*, 29–52. https://doi.org/10.1111/nyas.12360

Arenas, A., Díaz-Guilera, A., Kurths, J., Moreno, Y., & Zhou, C. (2008). Synchronization in complex networks. *Physics Reports*, *469*(3), 93–153. https://doi.org/10.1016/j.physrep.2008.09.002

Arnold, S. E., Ruscheinsky, D. D., & Han, L. Y. (1997). Further evidence of abnormal cytoarchitecture of the entorhinal cortex in schizophrenia using spatial point pattern analyses. *Biological Psychiatry*, *42*(8), 639–647. https://doi.org/10.1016/S0006-3223(97)00142-X

Bai, F., Watson, D. R., Yu, H., Shi, Y., Yuan, Y., & Zhang, Z. (2009). Abnormal resting-state functional connectivity of posterior cingulate cortex in amnestic type mild cognitive impairment. *Brain Research*, *1302*, 167–174. https://doi.org/10.1016/j.brainres.2009.09.028

Barabási, A.-L. (2016). *Network Science*. Cambridge University Press.

Betzel, R. F., Byrge, L., He, Y., Goñi, J., Zuo, X.-N., & Sporns, O. (2014). Changes in structural and functional connectivity among resting-state networks across the human lifespan. *Neuroimage*, *102 Pt 2*, 345–357. https://doi.org/10.1016/j.neuroimage.2014.07.067

Bharath, R. D., Munivenkatappa, A., Gohel, S., Panda, R., Saini, J., Rajeswaran, J., … Biswal, B. B. (2015). Recovery of resting brain connectivity ensuing mild traumatic brain injury. *Frontiers in Human Neuroscience*, *9*, 513. https://doi.org/10.3389/fnhum.2015.00513

Borich, M., Babul, A.-N., Yuan, P. H., Boyd, L., & Virji-Babul, N. (2015). Alterations in resting-state brain networks in concussed adolescent athletes. *Journal of Neurotrauma*, *32*(4), 265–271. https://doi.org/10.1089/neu.2013.3269

Breakspear, M. (2017). Dynamic models of large-scale brain activity. *Nature Neuroscience*, *20*(3), 340–352. https://doi.org/10.1038/nn.4497

Breakspear, M., Heitmann, S., & Daffertshofer, A. (2010). Generative models of cortical oscillations: neurobiological implications of the kuramoto model. *Frontiers in Human Neuroscience*, *4*, 190. https://doi.org/10.3389/fnhum.2010.00190

Buckner, R. L., Sepulcre, J., Talukdar, T., Krienen, F. M., Liu, H., Hedden, T., … Johnson, K. A. (2009). Cortical hubs revealed by intrinsic functional connectivity: mapping, assessment of stability, and relation to Alzheimer's disease. *The Journal of Neuroscience*, *29*(6), 1860–1873. https://doi.org/10.1523/JNEUROSCI.5062-08.2009

Buckner, R. L., Snyder, A. Z., Shannon, B. J., LaRossa, G., Sachs, R., Fotenos, A. F., … Mintun, M. A. (2005). Molecular, structural, and functional characterization of Alzheimer's disease: evidence for a relationship between default activity, amyloid, and memory. *The Journal of Neuroscience*, *25*(34), 7709–7717. https://doi.org/10.1523/JNEUROSCI.2177-05.2005

Cabral, J., Hugues, E., Kringelbach, M. L., & Deco, G. (2012). Modeling the outcome of structural disconnection on resting-state functional connectivity. *Neuroimage*, *62*(3), 1342–1353. https://doi.org/10.1016/j.neuroimage.2012.06.007

Cabral, J., Hugues, E., Sporns, O., & Deco, G. (2011). Role of local network oscillations in resting-state functional connectivity. *Neuroimage*, *57*(1), 130–139. https://doi.org/10.1016/j.neuroimage.2011.04.010

Cabral, J., Kringelbach, M. L., & Deco, G. (2014). Exploring the network dynamics underlying brain activity during rest. *Progress in Neurobiology*, *114*, 102–131. https://doi.org/10.1016/j.pneurobio.2013.12.005

Cabral, J., Kringelbach, M. L., & Deco, G. (2017). Functional connectivity dynamically evolves on multiple time-scales over a static structural connectome: Models and mechanisms. *Neuroimage*, *160*, 84–96. https://doi.org/10.1016/j.neuroimage.2017.03.045

Celone, K. A., Calhoun, V. D., Dickerson, B. C., Atri, A., Chua, E. F., Miller, S. L., … Sperling, R. A. (2006). Alterations in memory networks in mild cognitive impairment and Alzheimer's disease: an independent component analysis. *The Journal of Neuroscience*, *26*(40), 10222–10231. https://doi.org/10.1523/JNEUROSCI.2250-06.2006

Chang, C., & Glover, G. H. (2010). Time-frequency dynamics of resting-state brain connectivity measured with fMRI. *Neuroimage*, *50*(1), 81–98. https://doi.org/10.1016/j.neuroimage.2009.12.011

Chang, C.-C., Lee, Y. C., Lui, C.-C., & Lai, S.-L. (2007). Right anterior cingulate cortex infarction and transient speech aspontaneity. *Archives of Neurology*, *64*(3), 442–446. https://doi.org/10.1001/archneur.64.3.442

Corbetta, M., Kincade, J. M., & Shulman, G. L. (2002). Neural systems for visual orienting and their relationships to spatial working memory. *Journal of Cognitive Neuroscience*, *14*(3), 508–523. https://doi.org/10.1162/089892902317362029

de Reus, M. A., & van den Heuvel, M. P. (2013). Estimating false positives and negatives in brain networks. *Neuroimage*, *70*, 402–409. https://doi.org/10.1016/j.neuroimage.2012.12.066

Deco, G., Cabral, J., Saenger, V. M., Boly, M., Tagliazucchi, E., Laufs, H., … Kringelbach, M. L. (2018). Perturbation of whole-brain dynamics in silico reveals mechanistic differences between brain states. *Neuroimage*, *169*, 46–56. https://doi.org/10.1016/j.neuroimage.2017.12.009

Deco, G., & Jirsa, V. K. (2012). Ongoing cortical activity at rest: criticality, multistability, and ghost attractors. *The Journal of Neuroscience*, *32*(10), 3366–3375. https://doi.org/10.1523/JNEUROSCI.2523-11.2012

Deco, G., Jirsa, V. K., Robinson, P. A., Breakspear, M., & Friston, K. (2008). The dynamic brain: from spiking neurons to neural masses and cortical fields. *PLoS Computational Biology*, *4*(8), e1000092. https://doi.org/10.1371/journal.pcbi.1000092

Deco, G., Jirsa, V., McIntosh, A. R., Sporns, O., & Kötter, R. (2009). Key role of coupling, delay, and noise in resting brain fluctuations. *Proceedings of the National Academy of Sciences of the United States of America*, *106*(25), 10302–10307. https://doi.org/10.1073/pnas.0901831106

Deco, G., & Kringelbach, M. (2016). Metastability and Coherence: Extending the Communication through Coherence Hypothesis Using a Whole-Brain Computational Perspective. *Trends in Neurosciences*, *39*(6), 432. https://doi.org/10.1016/j.tins.2016.04.006

Deco, G., Kringelbach, M. L., Jirsa, V. K., & Ritter, P. (2017). The dynamics of resting fluctuations in the brain: metastability and its dynamical cortical core. *Scientific Reports*, *7*(1), 3095. https://doi.org/10.1038/s41598-017-03073-5

Deco, G., Van Hartevelt, T. J., Fernandes, H. M., Stevner, A., & Kringelbach, M. L. (2017). The most relevant human brain regions for functional connectivity: Evidence for a dynamical workspace of binding nodes from whole-brain computational

modelling. *Neuroimage*, *146*, 197–210. https://doi.org/10.1016/j.neuroimage.2016.10.047

Desikan, R. S., Ségonne, F., Fischl, B., Quinn, B. T., Dickerson, B. C., Blacker, D., … Killiany, R. J. (2006). An automated labeling system for subdividing the human cerebral cortex on MRI scans into gyral based regions of interest. *Neuroimage*, *31*(3), 968–980. https://doi.org/10.1016/j.neuroimage.2006.01.021

Fernández-Matarrubia, M., Matías-Guiu, J. A., Cabrera-Martín, M. N., Moreno-Ramos, T., Valles-Salgado, M., Carreras, J. L., & Matías-Guiu, J. (2018). Different apathy clinical profile and neural correlates in behavioral variant frontotemporal dementia and Alzheimer's disease. *International Journal of Geriatric Psychiatry*, *33*(1), 141–150. https://doi.org/10.1002/gps.4695

Fink, C. G. (2018). Resource Letter PB-1: The Physics of the Brain. *American Journal of Physics*, *86*(11), 805–817. https://doi.org/10.1119/1.5054288

Folke, C., Carpenter, S., Walker, B., Scheffer, M., Elmqvist, T., Gunderson, L., & Holling, C. S. (2004). REGIME SHIFTS, RESILIENCE, AND BIODIVERSITY IN ECOSYSTEM MANAGEMENT. *Annual Review of Ecology, Evolution, and Systematics*, *35*(1), 557–581. https://doi.org/10.1146/annurev.ecolsys.35.021103.105711

Fornito, A., Zalesky, A., & Bullmore, E. (2016). *Fundamentals of Brain Network Analysis*. Academic Press.

Fox, M. D., Corbetta, M., Snyder, A. Z., Vincent, J. L., & Raichle, M. E. (2006). Spontaneous neuronal activity distinguishes human dorsal and ventral attention systems. *Proceedings of the National Academy of Sciences of the United States of America*, *103*(26), 10046–10051. https://doi.org/10.1073/pnas.0604187103

Fransson, P., & Marrelec, G. (2008). The precuneus/posterior cingulate cortex plays a pivotal role in the default mode network: Evidence from a partial correlation network analysis. *Neuroimage*, *42*(3), 1178–1184. https://doi.org/10.1016/j.neuroimage.2008.05.059

Freeman, L. C. (1977). A Set of Measures of Centrality Based on Betweenness. *Sociometry*, *40*(1), 35. https://doi.org/10.2307/3033543

Freeman, L. C. (1978). Centrality in social networks conceptual clarification. *Social Networks*, *1*(3), 215–239. https://doi.org/10.1016/0378-8733(78)90021-7

Freyer, F., Roberts, J. A., Becker, R., Robinson, P. A., Ritter, P., & Breakspear, M. (2011). Biophysical mechanisms of multistability in resting-state cortical rhythms. *The Journal of Neuroscience*, *31*(17), 6353–6361. https://doi.org/10.1523/JNEUROSCI.6693-10.2011

Fukushima, M., Betzel, R. F., He, Y., de Reus, M. A., van den Heuvel, M. P., Zuo, X.-N., & Sporns, O. (2018). Fluctuations between high- and low-modularity topology in time-resolved functional connectivity. *Neuroimage*, *180*(Pt B), 406–416. https://doi.org/10.1016/j.neuroimage.2017.08.044

Fukushima, M., Betzel, R. F., He, Y., van den Heuvel, M. P., Zuo, X.-N., & Sporns, O. (2018). Structure-function relationships during segregated and integrated network states of human brain functional connectivity. *Brain Structure & Function*, *223*(3), 1091–1106. https://doi.org/10.1007/s00429-017-1539-3

Glasser, M. F., Sotiropoulos, S. N., Wilson, J. A., Coalson, T. S., Fischl, B., Andersson, J. L., … WU-Minn HCP Consortium. (2013). The minimal preprocessing pipelines for the Human Connectome Project. *Neuroimage*, *80*, 105–124. https://doi.org/10.1016/j.neuroimage.2013.04.127

Gollo, L. L., Roberts, J. A., Cropley, V. L., Di Biase, M. A., Pantelis, C., Zalesky, A., & Breakspear, M. (2018). Fragility and volatility of structural hubs in the human connectome. *Nature Neuroscience*, *21*(8), 1107–1116. https://doi.org/10.1038/s41593-018-0188-z

Gollo, L. L., Zalesky, A., Hutchison, R. M., van den Heuvel, M., & Breakspear, M. (2015). Dwelling quietly in the rich club: brain network determinants of slow cortical fluctuations. *Philosophical Transactions of the Royal Society of London. Series B, Biological Sciences*, *370*(1668). https://doi.org/10.1098/rstb.2014.0165

Greicius, M. D., Srivastava, G., Reiss, A. L., & Menon, V. (2004). Default-mode network activity distinguishes Alzheimer's disease from healthy aging: evidence from functional MRI. *Proceedings of the National Academy of Sciences of the United States of America*, *101*(13), 4637–4642. https://doi.org/10.1073/pnas.0308627101

Griffa, A., Baumann, P. S., Thiran, J.-P., & Hagmann, P. (2013). Structural connectomics in brain diseases. *Neuroimage*, *80*, 515–526. https://doi.org/10.1016/j.neuroimage.2013.04.056

Griffanti, L., Salimi-Khorshidi, G., Beckmann, C. F., Auerbach, E. J., Douaud, G., Sexton, C. E., … Smith, S. M. (2014). ICA-based artefact removal and accelerated fMRI acquisition for improved resting state network imaging. *Neuroimage*, *95*, 232–247. https://doi.org/10.1016/j.neuroimage.2014.03.034

Hagmann, P., Cammoun, L., Gigandet, X., Meuli, R., Honey, C. J., Wedeen, V. J., & Sporns, O. (2008). Mapping the structural core of human cerebral cortex. *PLoS Biology*, *6*(7), e159. https://doi.org/10.1371/journal.pbio.0060159

Hedden, T., Van Dijk, K. R. A., Becker, J. A., Mehta, A., Sperling, R. A., Johnson, K. A., & Buckner, R. L. (2009). Disruption of functional connectivity in clinically normal older adults harboring amyloid burden. *The Journal of Neuroscience*, *29*(40), 12686–12694. https://doi.org/10.1523/JNEUROSCI.3189-09.2009

Hilborn, R. C. (2000). *Chaos and nonlinear dynamics*. Oxford University Press. https://doi.org/10.1093/acprof:oso/9780198507239.001.0001

Hillary, F. G., Rajtmajer, S. M., Roman, C. A., Medaglia, J. D., Slocomb-Dluzen, J. E., Calhoun, V. D., … Wylie, G. R. (2014). The rich get richer: brain injury elicits hyperconnectivity in core subnetworks. *Plos One*, *9*(8), e104021. https://doi.org/10.1371/journal.pone.0104021

Hillary, F. G., Roman, C. A., Venkatesan, U., Rajtmajer, S. M., Bajo, R., & Castellanos, N. D. (2015). Hyperconnectivity is a fundamental response to neurological disruption. *Neuropsychology*, *29*(1), 59–75. https://doi.org/10.1037/neu0000110

Holling, C. S. (1973). Resilience and Stability of Ecological Systems. *Annual Review of Ecology and Systematics*, *4*(1), 1–23. https://doi.org/10.1146/annurev.es.04.110173.000245

Honey, C J, Sporns, O., Cammoun, L., Gigandet, X., Thiran, J. P., Meuli, R., & Hagmann, P. (2009). Predicting human resting-state functional connectivity from structural connectivity. *Proceedings of the National Academy of Sciences of the United States of America*, *106*(6), 2035–2040. https://doi.org/10.1073/pnas.0811168106

Honey, Christopher J, Kötter, R., Breakspear, M., & Sporns, O. (2007). Network structure of cerebral cortex shapes functional connectivity on multiple time scales. *Proceedings of the National Academy of Sciences of the United States of America*, *104*(24), 10240–10245. https://doi.org/10.1073/pnas.0701519104

Huijbers, W., Pennartz, C. M. A., Cabeza, R., & Daselaar, S. M. (2011). The hippocampus is coupled with the default network during memory retrieval but not during memory encoding. *Plos One*, *6*(4), e17463. https://doi.org/10.1371/journal.pone.0017463

Hutchison, R. M., Womelsdorf, T., Allen, E. A., Bandettini, P. A., Calhoun, V. D., Corbetta, M., … Chang, C. (2013). Dynamic functional connectivity: promise, issues, and interpretations. *Neuroimage*, *80*, 360–378. https://doi.org/10.1016/j.neuroimage.2013.05.079

Iraji, A., Chen, H., Wiseman, N., Welch, R. D., O'Neil, B. J., Haacke, E. M., … Kou, Z. (2016). Compensation through Functional Hyperconnectivity: A Longitudinal Connectome Assessment of Mild Traumatic Brain Injury. *Neural Plasticity*, *2016*, 4072402. https://doi.org/10.1155/2016/4072402

Japee, S., Holiday, K., Satyshur, M. D., Mukai, I., & Ungerleider, L. G. (2015). A role of right middle frontal gyrus in reorienting of attention: a case study. *Frontiers in Systems Neuroscience*, *9*, 23. https://doi.org/10.3389/fnsys.2015.00023

Kaboodvand, N., Bäckman, L., Nyberg, L., & Salami, A. (2018). The retrosplenial cortex: A memory gateway between the cortical default mode network and the medial temporal lobe. *Human Brain Mapping*, *39*(5), 2020–2034. https://doi.org/10.1002/hbm.23983

Kelso, J. A. S. (2008). An essay on understanding the mind. *Ecological Psychology : a Publication of the International Society for Ecological Psychology*, *20*(2), 180–208. https://doi.org/10.1080/10407410801949297

König, P., Engel, A. K., & Singer, W. (1996). Integrator or coincidence detector? The role of the cortical neuron revisited. *Trends in Neurosciences*, *19*(4), 130–137. https://doi.org/10.1016/S0166-2236(96)80019-1

Kuramoto, Y. (1984). *Chemical oscillations, waves, and turbulence* (Vol. 19). Berlin, Heidelberg: Springer Berlin Heidelberg. https://doi.org/10.1007/978-3-642-69689-3

Lavin, C., Melis, C., Mikulan, E., Gelormini, C., Huepe, D., & Ibañez, A. (2013). The anterior cingulate cortex: an integrative hub for human socially-driven interactions. *Frontiers in Neuroscience*, *7*, 64. https://doi.org/10.3389/fnins.2013.00064

Leech, R., Kamourieh, S., Beckmann, C. F., & Sharp, D. J. (2011). Fractionating the default mode network: distinct contributions of the ventral and dorsal posterior cingulate cortex to cognitive control. *The Journal of Neuroscience*, *31*(9), 3217–3224. https://doi.org/10.1523/JNEUROSCI.5626-10.2011

Leech, R., & Sharp, D. J. (2014). The role of the posterior cingulate cortex in cognition and disease. *Brain: A Journal of Neurology*, *137*(Pt 1), 12–32. https://doi.org/10.1093/brain/awt162

Mayer, A. R., Mannell, M. V., Ling, J., Gasparovic, C., & Yeo, R. A. (2011). Functional connectivity in mild traumatic brain injury. *Human Brain Mapping*, *32*(11), 1825–1835. https://doi.org/10.1002/hbm.21151

Meunier, D., Lambiotte, R., Fornito, A., Ersche, K. D., & Bullmore, E. T. (2009). Hierarchical modularity in human brain functional networks. *Frontiers in Neuroinformatics*, *3*, 37. https://doi.org/10.3389/neuro.11.037.2009

Mukamel, R., Gelbard, H., Arieli, A., Hasson, U., Fried, I., & Malach, R. (2005). Coupling between neuronal firing, field potentials, and FMRI in human auditory cortex. *Science*, *309*(5736), 951–954. https://doi.org/10.1126/science.1110913

Muller, A. M., & Virji-Babul, N. (2018). Stuck in a state of inattention? functional hyperconnectivity as an indicator of disturbed intrinsic brain dynamics in adolescents with concussion: A pilot study. *ASN Neuro*, *10*, 1759091417753802. https://doi.org/10.1177/1759091417753802

Nakamura, T., Hillary, F. G., & Biswal, B. B. (2009). Resting network plasticity following brain injury. *Plos One*, *4*(12), e8220. https://doi.org/10.1371/journal.pone.0008220

Newman, M. E. J. (2003). The Structure and Function of Complex Networks. *SIAM Review*, *45*, 167–256. https://doi.org/10.1137/S003614450342480

Omidvarnia, A., Pedersen, M., Walz, J. M., Vaughan, D. N., Abbott, D. F., & Jackson, G. D. (2016). Dynamic regional phase synchrony (DRePS): An Instantaneous Measure of Local fMRI Connectivity Within Spatially Clustered Brain Areas. *Human Brain Mapping*, *37*(5), 1970–1985. https://doi.org/10.1002/hbm.23151

Pedersen, M., Omidvarnia, A., Walz, J. M., Zalesky, A., & Jackson, G. D. (2017). Spontaneous brain network activity: Analysis of its temporal complexity. *Network Neuroscience (Cambridge, Mass.)*, *1*(2), 100–115. https://doi.org/10.1162/NETN_a_00006

Ponce-Alvarez, A., Deco, G., Hagmann, P., Romani, G. L., Mantini, D., & Corbetta, M. (2015). Resting-state temporal synchronization networks emerge from connectivity topology and heterogeneity. *PLoS Computational Biology*, *11*(2), e1004100. https://doi.org/10.1371/journal.pcbi.1004100

Power, J. D., Cohen, A. L., Nelson, S. M., Wig, G. S., Barnes, K. A., Church, J. A., … Petersen, S. E. (2011). Functional network organization of the human brain. *Neuron*, *72*(4), 665–678. https://doi.org/10.1016/j.neuron.2011.09.006

Purves, D., Augustine, G., Fitzpatrick, D., Hall, W., LaMantia, A., White, L., … Platt, M. (2004). *Neuroscience*. (D. Purves, G. Augustine, D. Fitzpatrick, W. Hall, A. LaMantia, L. White, … M. Platt, Eds.) (3rd ed.). Sunderland, MA 01375 U.S.A.: Oxford University Press.

Ries, A., Chang, C., Glim, S., Meng, C., Sorg, C., & Wohlschläger, A. (2018). Grading of Frequency Spectral Centroid Across Resting-State Networks. *Frontiers in Human Neuroscience*, *12*, 436. https://doi.org/10.3389/fnhum.2018.00436

Ritter, P., Schirner, M., McIntosh, A. R., & Jirsa, V. K. (2013). The virtual brain integrates computational modeling and multimodal neuroimaging. *Brain Connectivity*, *3*(2), 121–145. https://doi.org/10.1089/brain.2012.0120

Roberts, J. A., Friston, K. J., & Breakspear, M. (2017). Clinical applications of stochastic dynamic models of the brain, part I: A primer. *Biological Psychiatry : Cognitive

*Neuroscience and Neuroimaging*, *2*(3), 216–224.
https://doi.org/10.1016/j.bpsc.2017.01.010

Roberts, J. A., Gollo, L. L., Abeysuriya, R., Roberts, G., Mitchell, P. B., Woolrich, M. W., & Breakspear, M. (2018). Metastable brain waves. *BioRxiv*. https://doi.org/10.1101/347054

Rodrigues, F. A., Peron, T. K. D., Ji, P., & Kurths, J. (2016). The Kuramoto model in complex networks. *Physics Reports*, *610*, 1–98. https://doi.org/10.1016/j.physrep.2015.10.008

Röhm, A., Lüdge, K., & Schneider, I. (2018). Bistability in two simple symmetrically coupled oscillators with symmetry-broken amplitude- and phase-locking. *Chaos*, *28*(6), 063114. https://doi.org/10.1063/1.5018262

Rubinov, M., & Sporns, O. (2010). Complex network measures of brain connectivity: uses and interpretations. *Neuroimage*, *52*(3), 1059–1069. https://doi.org/10.1016/j.neuroimage.2009.10.003

Rubinov, M., & Sporns, O. (2011). Weight-conserving characterization of complex functional brain networks. *Neuroimage*, *56*(4), 2068–2079. https://doi.org/10.1016/j.neuroimage.2011.03.069

Saenger, V. M., Kahan, J., Foltynie, T., Friston, K., Aziz, T. Z., Green, A. L., … Deco, G. (2017). Uncovering the underlying mechanisms and whole-brain dynamics of deep brain stimulation for Parkinson's disease. *Scientific Reports*, *7*(1), 9882. https://doi.org/10.1038/s41598-017-10003-y

Salimi-Khorshidi, G., Douaud, G., Beckmann, C. F., Glasser, M. F., Griffanti, L., & Smith, S. M. (2014). Automatic denoising of functional MRI data: combining independent component analysis and hierarchical fusion of classifiers. *Neuroimage*, *90*, 449–468. https://doi.org/10.1016/j.neuroimage.2013.11.046

Salinas, E., & Sejnowski, T. J. (2001). Correlated neuronal activity and the flow of neural information. *Nature Reviews. Neuroscience*, *2*(8), 539–550. https://doi.org/10.1038/35086012

Scheffer, M., Carpenter, S., Foley, J. A., Folke, C., & Walker, B. (2001). Catastrophic shifts in ecosystems. *Nature*, *413*(6856), 591–596. https://doi.org/10.1038/35098000

Senden, M., Reuter, N., van den Heuvel, M. P., Goebel, R., & Deco, G. (2017). Cortical rich club regions can organize state-dependent functional network formation by engaging in oscillatory behavior. *Neuroimage*, *146*, 561–574. https://doi.org/10.1016/j.neuroimage.2016.10.044

Sharp, D. J., Beckmann, C. F., Greenwood, R., Kinnunen, K. M., Bonnelle, V., De Boissezon, X., … Leech, R. (2011). Default mode network functional and structural connectivity after traumatic brain injury. *Brain: A Journal of Neurology*, *134*(Pt 8), 2233–2247. https://doi.org/10.1093/brain/awr175

Shinoura, N., Yamada, R., Tabei, Y., Otani, R., Itoi, C., Saito, S., & Midorikawa, A. (2011). Damage to the right dorsal anterior cingulate cortex induces panic disorder. *Journal of Affective Disorders*, *133*(3), 569–572. https://doi.org/10.1016/j.jad.2011.04.029

Shumskaya, E., Andriessen, T. M. J. C., Norris, D. G., & Vos, P. E. (2012). Abnormal whole-brain functional networks in homogeneous acute mild traumatic brain injury. *Neurology*, *79*(2), 175–182. https://doi.org/10.1212/WNL.0b013e31825f04fb

Skardal, P. S., Ott, E., & Restrepo, J. G. (2011). Cluster synchrony in systems of coupled phase oscillators with higher-order coupling. *Physical Review. E, Statistical, Nonlinear, and Soft Matter Physics*, *84*(3 Pt 2), 036208. https://doi.org/10.1103/PhysRevE.84.036208

Smith, S. M., Beckmann, C. F., Andersson, J., Auerbach, E. J., Bijsterbosch, J., Douaud, G., … WU-Minn HCP Consortium. (2013). Resting-state fMRI in the Human Connectome Project. *Neuroimage*, *80*, 144–168. https://doi.org/10.1016/j.neuroimage.2013.05.039

Smith, S. M., Fox, P. T., Miller, K. L., Glahn, D. C., Fox, P. M., Mackay, C. E., … Beckmann, C. F. (2009). Correspondence of the brain's functional architecture during activation and rest. *Proceedings of the National Academy of Sciences of the United States of America*, *106*(31), 13040–13045. https://doi.org/10.1073/pnas.0905267106

Sporns, O., Honey, C. J., & Kötter, R. (2007). Identification and classification of hubs in brain networks. *Plos One*, *2*(10), e1049. https://doi.org/10.1371/journal.pone.0001049

Spreng, R. N., Sepulcre, J., Turner, G. R., Stevens, W. D., & Schacter, D. L. (2013). Intrinsic architecture underlying the relations among the default, dorsal attention, and frontoparietal control networks of the human brain. *Journal of Cognitive Neuroscience*, *25*(1), 74–86. https://doi.org/10.1162/jocn_a_00281

Strogatz, S H. (2001). Exploring complex networks. *Nature*, *410*, 268–276. https://doi.org/10.1038/35065725

Strogatz, Steven H. (2018). *Nonlinear dynamics and chaos: with applications to physics, biology, chemistry, and engineering*. CRC Press. https://doi.org/10.1201/9780429492563

Thompson, W. H., Brantefors, P., & Fransson, P. (2017). From static to temporal network theory: Applications to functional brain connectivity. *Network Neuroscience (Cambridge, Mass.)*, *1*(2), 69–99. https://doi.org/10.1162/NETN_a_00011

Tian, L., Jiang, T., Wang, Y., Zang, Y., He, Y., Liang, M., … Zhuo, Y. (2006). Altered resting-state functional connectivity patterns of anterior cingulate cortex in adolescents with attention deficit hyperactivity disorder. *Neuroscience Letters*, *400*(1-2), 39–43. https://doi.org/10.1016/j.neulet.2006.02.022

Tisserand, D. J., Visser, P. J., van Boxtel, M. P., & Jolles, J. (2000). The relation between global and limbic brain volumes on MRI and cognitive performance in healthy individuals across the age range. *Neurobiology of Aging*, *21*(4), 569–576.

Tognoli, E., & Kelso, J. A. S. (2014). The metastable brain. *Neuron*, *81*(1), 35–48. https://doi.org/10.1016/j.neuron.2013.12.022

Tomasi, D., & Volkow, N. D. (2010). Functional connectivity density mapping. *Proceedings of the National Academy of Sciences of the United States of America*, *107*(21), 9885–9890. https://doi.org/10.1073/pnas.1001414107

Van Essen, D. C., Ugurbil, K., Auerbach, E., Barch, D., Behrens, T. E. J., Bucholz, R., … WU-Minn HCP Consortium. (2012). The Human Connectome Project: a data acquisition perspective. *Neuroimage*, *62*(4), 2222–2231. https://doi.org/10.1016/j.neuroimage.2012.02.018

Van Hoesen, G. W., Hyman, B. T., & Damasio, A. R. (1991). Entorhinal cortex pathology in Alzheimer's disease. *Hippocampus*, *1*(1), 1–8. https://doi.org/10.1002/hipo.450010102

van den Heuvel, M. P., de Reus, M. A., Feldman Barrett, L., Scholtens, L. H., Coopmans, F. M. T., Schmidt, R., … Li, L. (2015). Comparison of diffusion tractography and tract-tracing measures of connectivity strength in rhesus macaque connectome. *Human Brain Mapping*, *36*(8), 3064–3075. https://doi.org/10.1002/hbm.22828

van den Heuvel, M. P., Kahn, R. S., Goñi, J., & Sporns, O. (2012). High-cost, high-capacity backbone for global brain communication. *Proceedings of the National Academy of Sciences of the United States of America*, *109*(28), 11372–11377. https://doi.org/10.1073/pnas.1203593109

van den Heuvel, M. P., & Sporns, O. (2011). Rich-club organization of the human connectome. *The Journal of Neuroscience*, *31*(44), 15775–15786. https://doi.org/10.1523/JNEUROSCI.3539-11.2011

van den Heuvel, M. P., & Sporns, O. (2013). Network hubs in the human brain. *Trends in Cognitive Sciences*, *17*(12), 683–696. https://doi.org/10.1016/j.tics.2013.09.012

Vannini, P., O'Brien, J., O'Keefe, K., Pihlajamäki, M., Laviolette, P., & Sperling, R. A. (2011). What goes down must come up: role of the posteromedial cortices in encoding and retrieval. *Cerebral Cortex*, *21*(1), 22–34. https://doi.org/10.1093/cercor/bhq051

Váša, F., Shanahan, M., Hellyer, P. J., Scott, G., Cabral, J., & Leech, R. (2015). Effects of lesions on synchrony and metastability in cortical networks. *Neuroimage*, *118*, 456–467. https://doi.org/10.1016/j.neuroimage.2015.05.042

Ward, A. M., Mormino, E. C., Huijbers, W., Schultz, A. P., Hedden, T., & Sperling, R. A. (2015). Relationships between default-mode network connectivity, medial temporal lobe structure, and age-related memory deficits. *Neurobiology of Aging*, *36*(1), 265–272. https://doi.org/10.1016/j.neurobiolaging.2014.06.028

Weerasinghe, G., Duchet, B., Cagnan, H., Brown, P., Bick, C., & Bogacz, R. (2018). Predicting the effects of deep brain stimulation using a reduced coupled oscillator model. *BioRxiv*. https://doi.org/10.1101/448290

Yaesoubi, M., Allen, E. A., Miller, R. L., & Calhoun, V. D. (2015). Dynamic coherence analysis of resting fMRI data to jointly capture state-based phase, frequency, and time-domain information. *Neuroimage*, *120*, 133–142. https://doi.org/10.1016/j.neuroimage.2015.07.002

Yan, C.-G., Craddock, R. C., Zuo, X.-N., Zang, Y.-F., & Milham, M. P. (2013). Standardizing the intrinsic brain: towards robust measurement of inter-individual variation in 1000 functional connectomes. *Neuroimage*, *80*, 246–262. https://doi.org/10.1016/j.neuroimage.2013.04.081

Yan, H., Tian, L., Yan, J., Sun, W., Liu, Q., Zhang, Y.-B., … Zhang, D. (2012). Functional and anatomical connectivity abnormalities in cognitive division of anterior cingulate cortex in schizophrenia. *Plos One*, *7*(9), e45659. https://doi.org/10.1371/journal.pone.0045659

Yeh, F.-C., Wedeen, V. J., & Tseng, W.-Y. I. (2010). Generalized q-sampling imaging. *IEEE Transactions on Medical Imaging*, *29*(9), 1626–1635. https://doi.org/10.1109/TMI.2010.2045126

Yeo, B. T. T., Krienen, F. M., Sepulcre, J., Sabuncu, M. R., Lashkari, D., Hollinshead, M., … Buckner, R. L. (2011). The organization of the human cerebral cortex estimated by intrinsic functional connectivity. *Journal of Neurophysiology*, *106*(3), 1125–1165. https://doi.org/10.1152/jn.00338.2011

Young, C. K., & McNaughton, N. (2009). Coupling of theta oscillations between anterior and posterior midline cortex and with the hippocampus in freely behaving rats. *Cerebral Cortex*, *19*(1), 24–40. https://doi.org/10.1093/cercor/bhn055

Zalesky, A., Fornito, A., Cocchi, L., Gollo, L. L., & Breakspear, M. (2014). Time-resolved resting-state brain networks. *Proceedings of the National Academy of Sciences of the United States of America*, *111*(28), 10341–10346. https://doi.org/10.1073/pnas.1400181111

Zuo, X.-N., Ehmke, R., Mennes, M., Imperati, D., Castellanos, F. X., Sporns, O., & Milham, M. P. (2012). Network centrality in the human functional connectome. *Cerebral Cortex*, *22*(8), 1862–1875. https://doi.org/10.1093/cercor/bhr269

**Supplementary Material**

***Supplementary Equations.*** *The optimal working point of the model was estimated using a composite distance score which was calculated as the average of five unity-based normalized distance measures, as below:*

$$Composite\ Distance\ Score = (|UN(synchrony_{Sim} - \overline{synchrony_{emp}})| + |UN(metastability_{Sim} - \overline{metastability_{emp}})| + |UN(KSD)| + |1 - UN(Q)| + |1 - UN(CORR)|)/5$$

$$UN(x) = \frac{x - \min(x)}{\max(x) - \min(x)}$$

*$*UN$, Unity normalization; $KSD$, Kolmogorov-Smirnoff distance between the similarities of coherence measures, obtained from the empirical BOLD data and simulated signals; $CORR$, Pearson correlation coefficient between the empirical FC matrix and the FC matrix obtained for the simulated signals; $Q$, Whole brain modularity.

***Supplementary Table 1.*** *The list of abbreviations and complete names for all 68 regions-of-interest as included in the 'Desikan-Killiany' cortical atlas are shown in the table.*

| Abbreviation | Complete Name |
| --- | --- |
| BSTS | Banks superior temporal sulcus |
| CAC | Caudal anterior-cingulate cortex |
| CMF | Caudal middle frontal gyrus |
| CUN | Cuneus cortex |
| ENT | Entorhinal cortex |
| FUS | Fusiform gyrus |
| INFP | Inferior parietal cortex |
| IT | Inferior temporal gyrus |
| ISTC | Isthmus – cingulate cortex |
| LOCC | Lateral occipital cortex |
| LORB | Lateral orbital frontal cortex |
| LIN | Lingual gyrus |
| MORB | Medial orbital frontal cortex |
| MT | Middle temporal gyrus |
| PARH | Parahippocampal gyrus |
| PARC | Paracentral lobule |
| POPE | Pars opercularis |
| PORB | Pars orbitalis |
| PTRI | Pars triangularis |
| PCAL | Pericalcarine cortex |
| PSTS | Postcentral gyrus |
| PC | Posterior-cingulate cortex |
| PREC | Precentral gyrus |
| PCUN | Precuneus cortex |
| RAC | Rostral anterior cingulate cortex |
| RMF | Rostral middle frontal gyrus |
| SF | Superior frontal gyrus |
| SP | Superior parietal cortex |
| ST | Superior temporal gyrus |
| SMAR | Supramarginal gyrus |
| FP | Frontal pole |
| TP | Temporal pole |
| TT | Transverse temporal cortex |
| INS | Insular cortex |

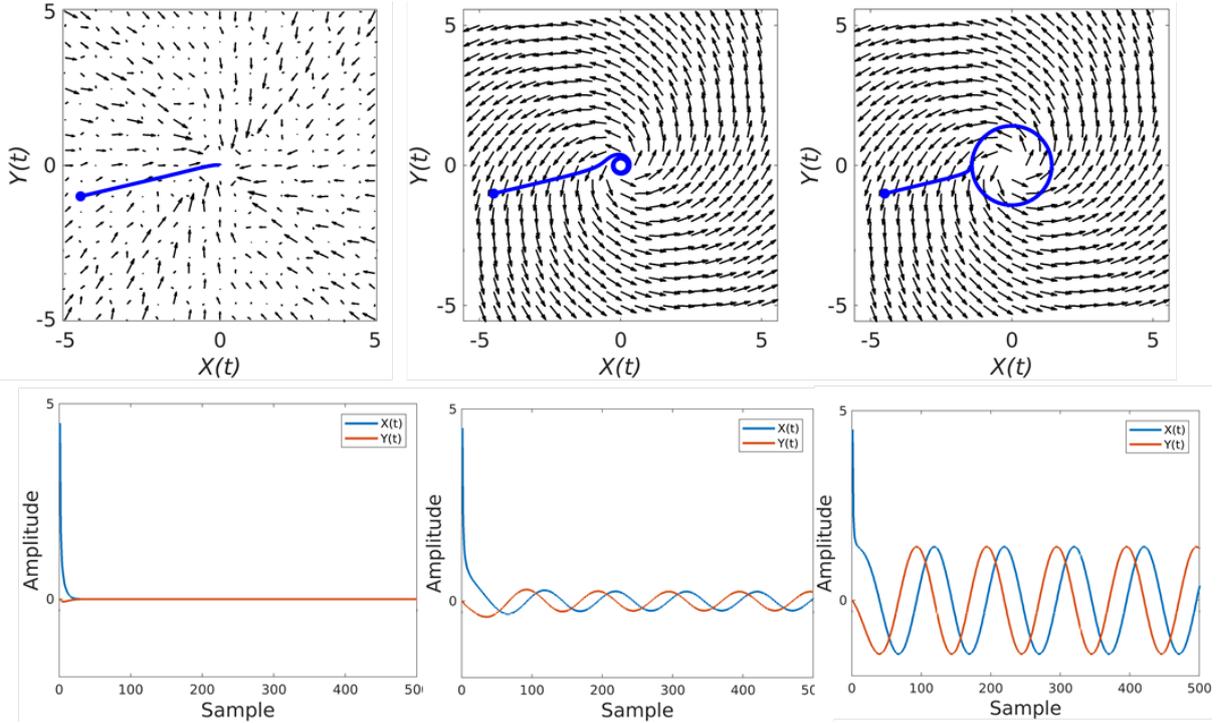

*Supplementary Figure 1. An illustrative example of phase space vector fields as well as temporal evolution of the Stuart-Landau oscillator. where $z = r\,e^{i\theta} = x + iy$ is a complex number describing the state of the oscillator, $\omega \in \mathbb{R}$ is the frequency of each oscillator and the bifurcation parameter $a \in \mathbb{R}$ determines whether the oscillator is characterized by noisy fluctuations or exhibits oscillatory behavior. The origin (i.e., $z = 0$) is the fixed point of this system. The eigenvalues of this system all have complex conjugate values ($\sigma = a \pm i\omega$), indicating a spiral trajectory behavior in the neighborhood of the fixed point in phase space. If $a < 0$, then the origin is a stable equilibrium solution with solutions spiraling into the origin (as illustrated in the first column of the figure). However, if $a > 0$, then the origin is an unstable equilibrium with solutions spiraling out from the origin (as illustrated in the third column of the figure). The illustrated closed orbit in the phase space represents the periodic behavior of the system. Solutions that reside inside of the closed orbit will spiral out towards the orbit, while solutions outside of the orbit will spiral inward. The middle column includes an example phase space and the associated signals at the bifurcation point ($a = 0$).*

# Resting State Networks

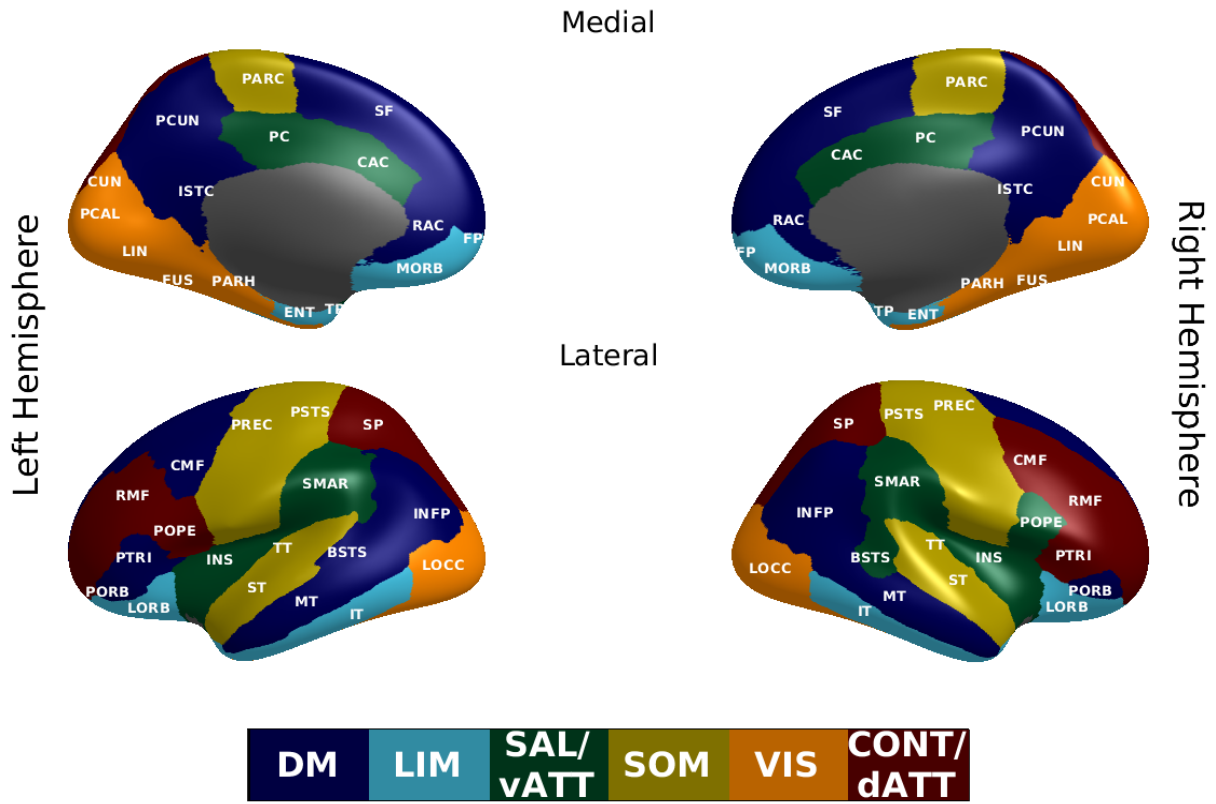

***Supplementary Figure 2.*** *The layout of the functional networks. The used abbreviations for brain regions are shown in Supplementary Table 1. DM, default mode network (DM); LIM, the limbic system (LIM); dATT/CONT, the dorsal attention or control network; SAL/vATT, the salience or ventral attention network; SOM, the somatomotor network; VIS, the visual network.*

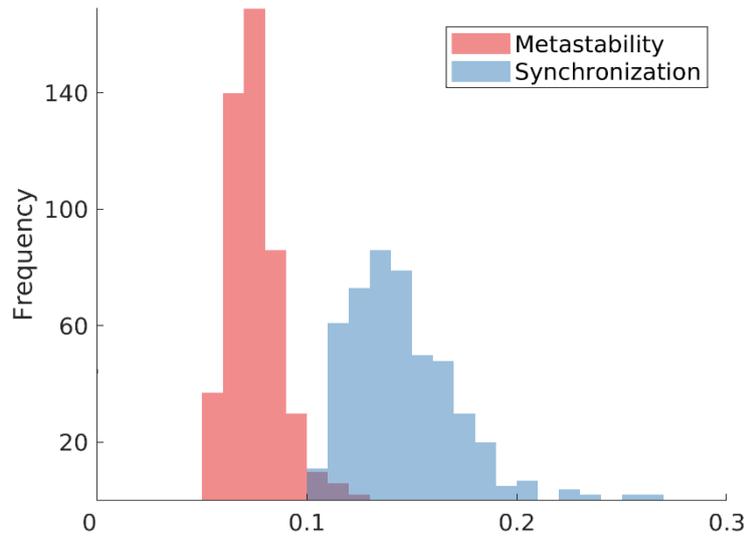

***Supplementary Figure 3.*** *Distribution of global synchrony and global metastability as indicative of macroscopic coherence of the whole-brain network are illustrated for the empirical BOLD signals.*

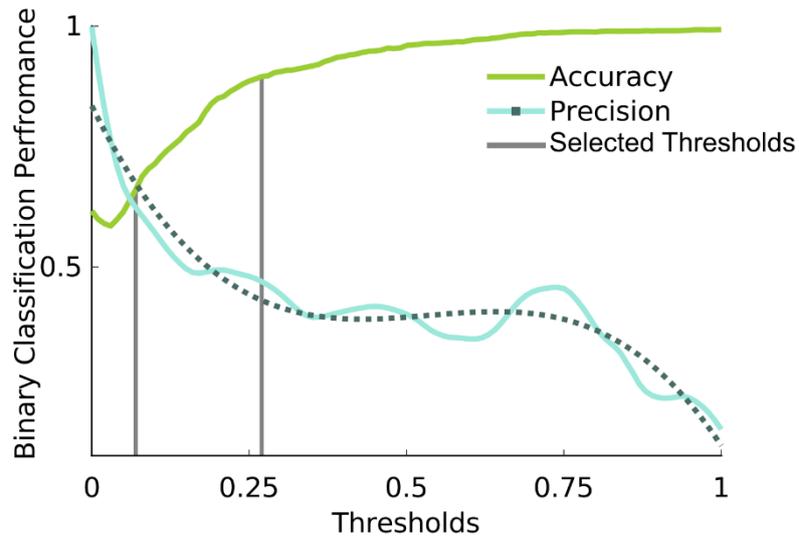

***Supplementary Figure 4.*** *The accuracy and precision of the binary classification of edges at different thresholds ranging from 0 to 1 are illustrated. Cubic polynomial curve fitted to the precision is depicted as the dotted green curve. The locations of the intersection of two performance measures, as well as the knee point for both curves refer to the thresholds of 0.08 and 0.27, respectively.*

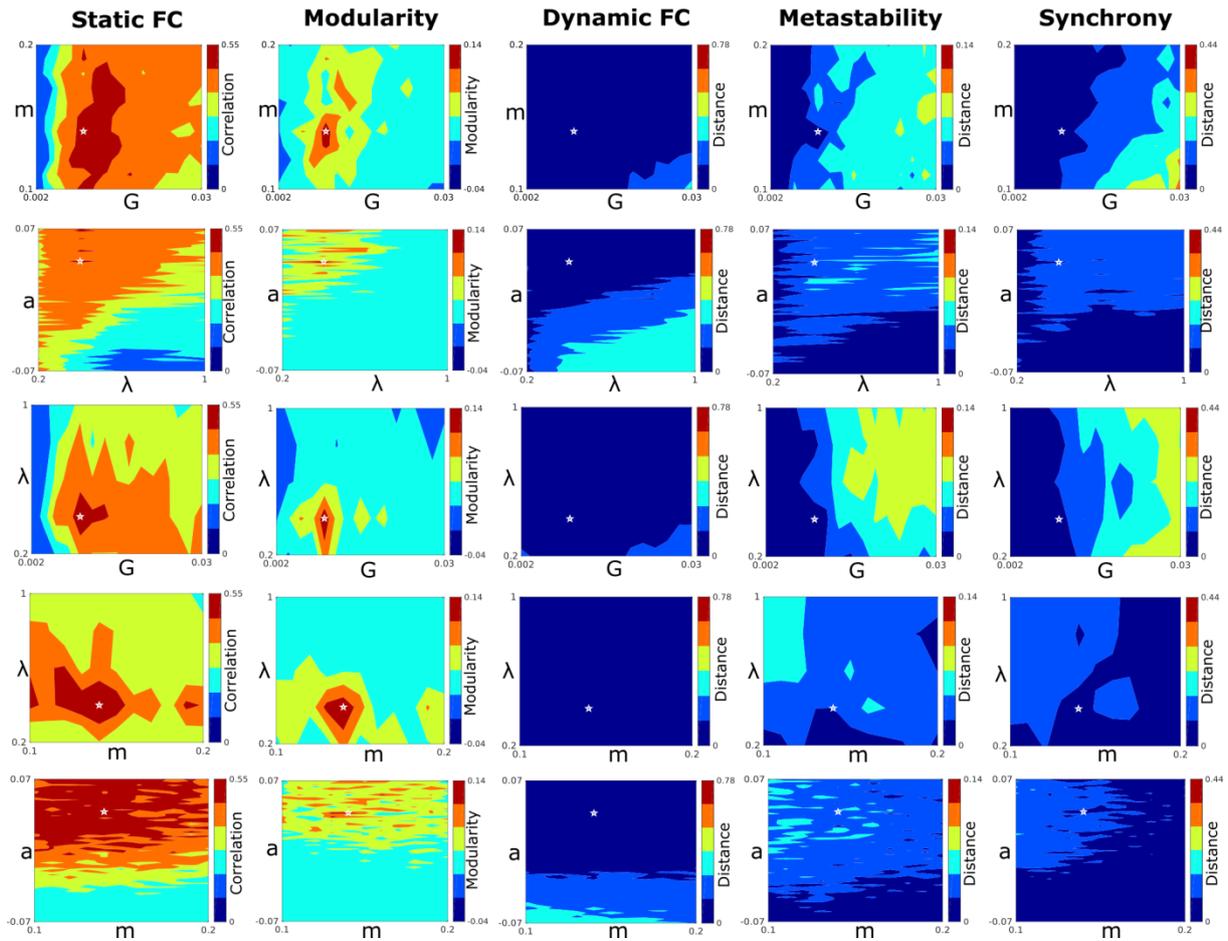

***Supplementary Figure 5.*** *Parameter space search. This figure shows the exploration of the parameter space defined by the bifurcation parameter a, global coupling G, frequency lethargy λ and frequency modulation m. First column depicts the Pearson correlation between empirical and simulated static FC patterns for different pairings of parameters. Second column shows the whole brain modularity computed for the simulated static FC matrix. The Kolmogorov-Smirnoff distance between the similarities of coherence measures, obtained from the empirical BOLD data and simulated signals, as well as the difference of metastability and synchrony of simulated signals from the average metastability and synchrony measures of empirical BOLD signals are respectively illustrated in columns 3-5. Measures associated with the optimal choice of global coupling (G), bifurcation parameter (a), frequency lethargy λ and frequency modulation m are shown as a white asterisk (G=0.01, a=0.038, λ=0.4 and m=0.14).*

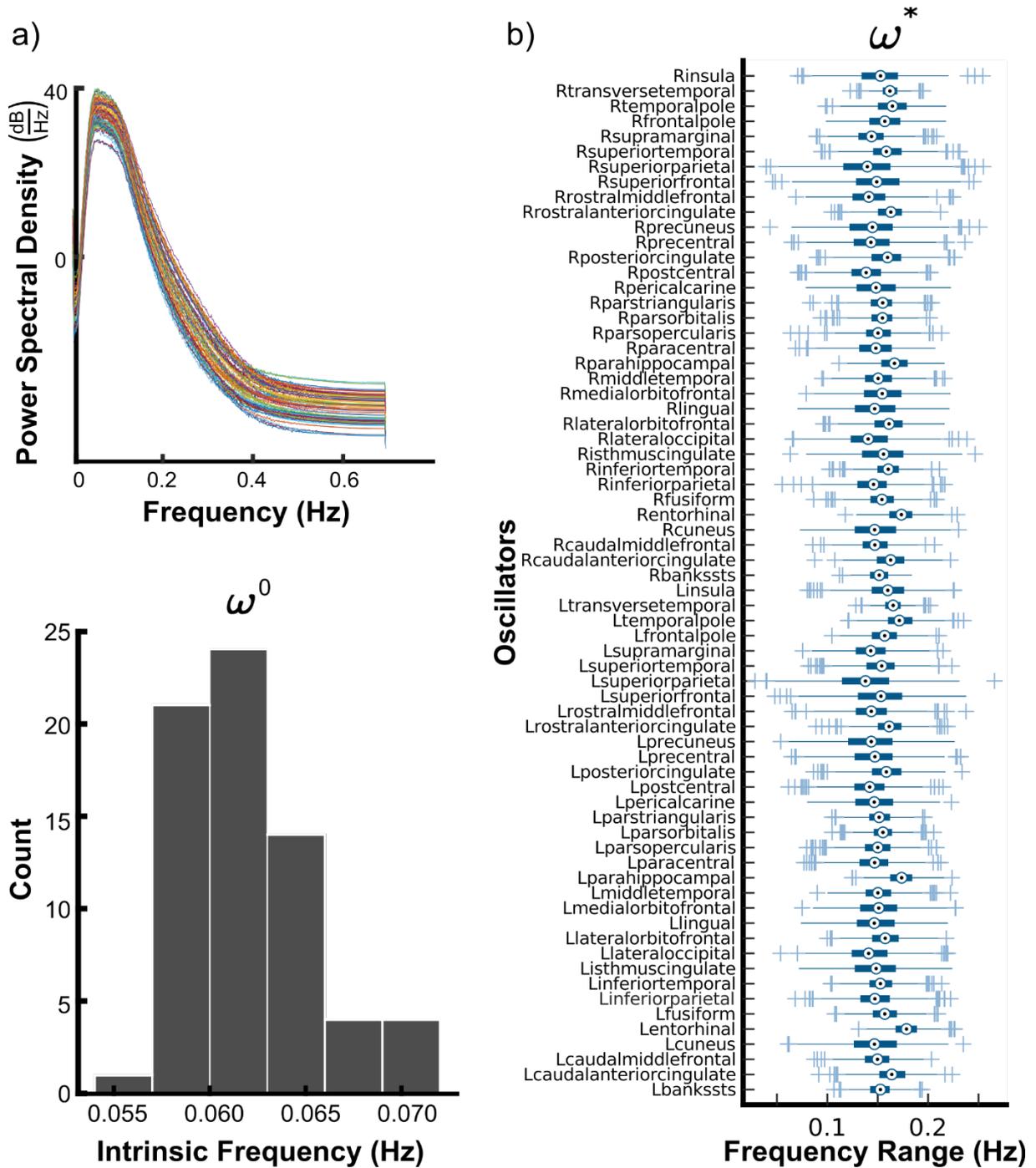

***Supplementary Figure 6.*** *(a) The power spectral density and distribution of regional intrinsic frequencies, calculated as the median (across subjects) peak frequency of the regional BOLD signals. (b) The working frequency range of oscillators, centered at $\boldsymbol{\omega}^* = 2.5\,\boldsymbol{\omega}^0$.*